# Understanding and Designing the Spin-Driven Thermoelectrics[*]

Md Mobarak Hossain Polash,[1,2] Duncan Moseley,[3] Junjie Zhang,[3,†] Raphaël P. Hermann,[3] Daryoosh Vashaee[1,2,‡§]

[1]Department of Materials Science and Engineering, NC State University, Raleigh, NC 27606, USA
[2]Department of Electrical and Computer Engineering, NC State University, Raleigh, NC 27606, USA
[3]Materials Science and Technology Division, Oak Ridge National Laboratory, Oak Ridge, TN 37831, USA

**Summary**

While the thermoelectric materials progress based on the engineering of electronic and phononic characteristics is reaching a plateau, adding the spin degree of freedom has the potential to open a new landscape for alternative thermoelectric materials. Here we present the concepts, current understanding, and guidelines for designing spin-driven thermoelectrics. We show that the interplay between the spin and heat currents in entropy transport via charge carriers can offer a strategic path to enhance the electronic thermopower. The classical antiferromagnetic semiconductor manganese telluride (MnTe) is chosen as the case study due to its significant spin-mediated thermoelectric properties. We show that although the spin-disorder scattering reduces the carrier mobility in magnetic materials, spin entropy, magnon, and paramagnon carrier drags can dominate over and significantly enhance the thermoelectric power factor and hence zT. Finally, several guidelines are drawn based on the current understandings for designing high-performance spin-driven thermoelectric materials.

**Introduction**

With the progress in developing high-efficiency thermoelectric materials over the last two decades, the thermoelectric market continues to grow for various applications, including power generation, cooling, sensing, and imaging applications. Accordingly, new generations of thermoelectric devices [1-5] and characterization techniques [6-8] have also being developed. New TE materials have been designed based on original concepts such as power factor enhancement via carrier filtering [1,9], carrier pocket engineering [10-12], complex structures [13,14], creation of resonant energy levels close to the band edges [15], and low dimensional structures [16,17], as well as enhanced phonon scattering via nano inclusions [18-20] and nanostructuring [21,22], have been implemented to improve the energy conversion efficiencies of the TE materials. In addition, new materials based on nano bulk forms such as nanocomposites and nanostructured single component bulk materials have especially taken attention due to their ease of fabrication and compatibility with the existing form of the TE devices [23,24]. Although nanostructuring methods were proven to be beneficial in many material systems [25,26], due to the interdependency of the thermal and electrical transport parameters, these methods are frequently associated with deterioration of the carrier mobility that in some materials can result in a significant reduction of the power factor leading to smaller figure-of-merit [27-29].

---

[*]This paper has been co-authored by employees of UT-Battelle, LLC under Contract No. DE-AC05-00OR22725 with the U.S. Department of Energy. The U.S. Government retains and the publisher, by accepting the paper for publication, acknowledges that the U.S. Government retains a nonexclusive, paid-up, irrevocable, worldwide license to publish or reproduce the published form of this paper, or allow others to do so, for U.S. Government purposes.
[†] Present address: State Key Laboratory of Crystal Materials, Shandong University, Jinan, Shandong 250100, China
[‡] Lead Contact
[§] Corresponding Author: dvashae@ncsu.edu

Despite the persistent ongoing efforts and investments, developing good thermoelectric materials based on the engineering of the electron and phonon transport properties is reaching a plateau. In recent years, magnetic semiconductors have been introduced to offer a new landscape to engineer thermoelectric materials based on the spin's contributions. The motivation originates from the fact that magnons (spin-wave) are bosons and are not bound to the fermionic limitation that introduces a counter-indicative relation between electrical conductivity and thermopower through the Fermi energy [30-31]. Magnetic materials can deliver spin contributions to the thermopower over a broad range of temperatures from below to above room temperatures. Some transition metal oxides have also shown a spin contribution to thermopower from crystal-field driven spin entropy [32-34]. As such, they can offer a prospect for various thermoelectric technologies, including energy harvesting, sensing, or cooling applications [35-37] with a high performance coming from the auspicious spin-caloritronic effects [37-41]. A complete understanding of the thermoelectric spin-based impact in magnetic semiconductors can pave the way to engineer the thermoelectric materials with larger power factors beyond the fermionic limitations [30-31]. In spin-caloritronic systems, spin can exist as the individual spin of itinerant carriers, spin-wave, and spin-wave packets. Spin-wave or spin-wave packets, the localized coupled spin ensembles, a.k.a. magnons, and paramagnons, respectively, behave as bosonic quasiparticles in the condensed matter [37,39]. The addition of the spin degree of freedom into the linear Onsager system having reciprocally coupled charge and phonon [42,43], can offer an excess thermopower contribution to the diffusion one, i.e., the entropy carried by free electrons, $\alpha_e = (\pi k_B)^2 T / 3eE_F$ [30-31]. Over the decades, various spin-caloritronic effects on thermopower have been studied, such as spin fluctuation systems [44-45], heavy fermion effects in Kondo lattices [46-48], dilute Kondo systems [49-51], Spin-Seebeck and Spin-Peltier effects [52-53], Spin-dependent Seebeck and Peltier effects [54-55], Spin entropy in hopping systems [32-34, 40], magnon-electron drag effect [37,39], paramagnon-electron drag effect [37,39], and magnon-bipolar carrier drag effect [31], but until now none of these effects have led towards a high thermoelectric figure-of-merit dominantly due to the spin effects. Table 1 summarized different material systems studied so far, showing spin-driven thermoelectric properties.

Table 1: Selected material studies demonstrating spin-driven thermoelectric properties

| Spin-caloritronic Effects | Materials | Impacts | Ref |
|---|---|---|---|
| Magnon-carrier drag | Fe, Co, Ni, MnTe | Theoretical view of the magnon-carrier drag | 56,57,58 |
| Spin-fluctuation | $Fe_2VAl_{0.9}Si_{0.1}$, $AT_4Sb_{12}$ skutterudites | Thermopower enhancement | 59,60 |
| Spin-entropy (spin and configurational degeneracies) | BiCuSeO, $Na_xCo_2O_4$ | Thermopower enhancement from spin entropy | 40,61 |
| Spin-Seebeck | Pt/Holey $MoS_2$/$Y_3Fe_5O_{12}$, Bi:YIG film, $NiFe_2O_4$/Pt | Thermopower enhancement | 62,63,64 |
| Half-metallicity | Heusler and half-heusler families | High zT from engineering of electron and phonon transport properties | 65,66 |
| Superparamagnetism | $x$TM/$Ba_{0.3}In_{0.3}Co_4Sb_{12}$ (TM = Co, Fe or Ni), $Bi_{0.5}Sb_{1.5}Te_3$ with $Fe_3O_4$ nanoparticles | Enhanced thermoelectric performance from magnetic and superparamagnetic fluctuations, zT≈1.8, 32% enhancement due to spins | 67,68 |

| Spin thermodynamic entropy | GeMnTe$_2$ | zT≈1.4, ~45% enhancement due to spins and carrier concentration optimization | 69 |
| Paramagnon-carrier drag | MnTe, Mn$_{1-x}$Cr$_x$Sb | Thermopower, zT≈1, ~300% enhancement due to spins | 37,39 |

The base materials for most of the systems listed in table 1 are already good thermoelectric materials where some magnetic doping is used to further increase the already high zT. The scenario is different for the manganese telluride (MnTe) system, a simple binary antiferromagnetic (AFM) semiconductor with a hexagonal NiAs crystal structure. The spin effects like magnon/paramagnon-drag and spin-disorder scattering have been observed in many ferromagnetic (FM) and AFM materials with similar trends as in MnTe [37,39,56,57,70]. MnTe displays spin effects greatly more significant than electronic effects on thermoelectric properties leading to zT≈1. Without the spin effects, MnTe is not a good thermoelectric material (electronic diffusion transport at optimum carrier concentration gives zT≈0.3). However, zT enhances by ~300% solely due to the spin effects. This enhancement is observed, interestingly, in the paramagnetic domain, where the magnetic ordering is diminished. Therefore, we base this study on the MnTe system to emphasize the spin-driven effects.

Recent studies on MnTe have demonstrated the significant spin-based thermopower contribution from paramagnon-hole drag, leading to zT > 1 [37] in the deep paramagnetic domain. The salient feature of paramagnon-carrier drag compared to other spin-caloritronic effects is that paramagnons are originated from the long spin-spin correlation that lived in the short-range ordered domains above the transition temperature. Moreover, MnTe is a simple binary spin system that can provide better insight into designing high-performance spin-driven thermoelectrics compared to the complex materials where synergistic electronic and spin effects cause a high zT. To avoid any confusion, we like to clarify that in this article, we are not reporting a new material system; instead, our focus is to develop a guideline for designing high-performance spin-driven thermoelectric materials. MnTe is chosen for the abovementioned reasons; furthermore, all the needed data for this study are available to the authors. However, for self-consistency, we resynthesized and characterized similar samples as in reference 37. Furthermore, we reproduced the reported data and performed further characterizations to understand and develop a generalized guideline for the broad thermoelectric community.

In FM and AFM materials, the coupling between phonons and magnons due to the thermal fluctuations of the localized spins' long-range ordered structure induces a momentum gradient into the magnon system, which can influence the itinerate electrons or holes via *s-d* or *p-d* interaction [37,39]. This transfer of linear momentum creates a drag effect on the electrons or holes. Hence, an advective transport mechanism is introduced to the charge carriers resulting in the excess contribution of magnon-drag thermopower to the diffusion thermopower. Figure 1(a) illustrates the magnon-electron drag (MED) mechanism schematically into a magnetically ordered material system.

The paramagnon-drag may occur above the transition temperature in short to mid-range ordered magnetic structures. A short-lived spin-wave in short to mid-range ordered structures can still act as a wave packet or quasi-magnon (paramagnon) due to the local thermal fluctuations of magnetization [37,39]. These paramagnons in the PM regime are expected to drag electrons (or holes) when the paramagnon lifetime is greater than the charge carrier momentum relaxation time, and the spatial spin-spin correlation length is larger than the mean free path and effective Bohr radius of the electrons [37].

On the other hand, spin disorder scattering theory has been widely used to explain the carrier mobility in magnetic materials. The *s-d* exchange interaction plays a vital role in both spin disorder scattering and magnon drag theories. However, theoretical calculations of the thermoelectric properties based on the existing theories cannot adequately explain the experimental data. AFM MnTe, a widely studied magnetic system, and a promising thermoelectric compound [37,[71]-[73]], makes an informative platform to explore the underlying physical phenomena related to the thermoelectric properties.

This study will discuss the fundamentals, prospects, and limitations of some of the most critical spin-based theories to explain the thermoelectric properties. We will choose MnTe for the case study to apply the theoretical analysis. The experimental results of undoped and Li-doped MnTe are considered to benchmark the theoretical predictions. We also evaluate the prospect of spin entropy effects to explain the excess thermopower of the MnTe system in the paramagnetic domain. Moreover, we calculate various magnetic heat capacity contributions of MnTe relevant to the magnon-drag thermopower. Such detailed discussions offer a better understanding of the existing theories and help devise more accurate formalisms for designing spin-based thermoelectric materials.

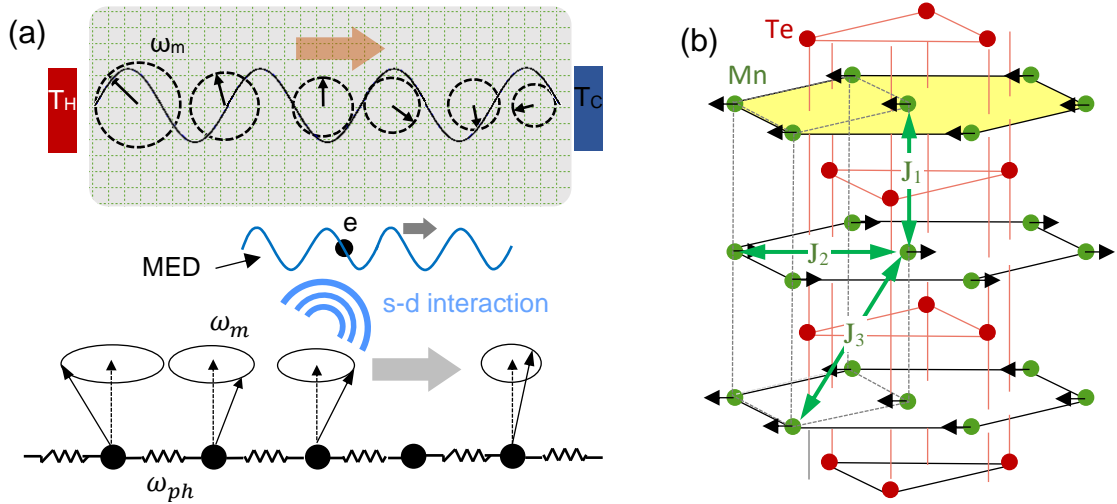

Figure 1: (a) Schematic demonstration of the magnon-electron drag (MED) in a magnetic material. $T_H$ and $T_C$ are the temperatures at hot and cold ends, $\omega_m$ and $\omega_{ph}$ are the magnon and phonon frequency, and *e* represents an electron. (b) the crystal of α-MnTe in the hexagonal NiAs structure. $J_1$, $J_2$, and $J_3$ represent the magnetic exchanges between pairs of first, second, and third nearest neighbor Mn ions, respectively.

## Results and Discussion

### Thermoelectric Transport Properties of MnTe and $Mn_xLi_{1-x}Te$

The AFM MnTe has a hexagonal NiAs structure [74], as shown in Figure 1(b). In MnTe, $Mn^{2+}$ ($3d^5$) ions have a $^6S$ ground spin state with the orbital angular momentum of $L = 0$ and spin angular momentum of $S = 5/2$ [75,76]. Due to the quenched orbital angular moment, the spin angular moment is the origin of the total magnetic moment of MnTe. Therefore, the terms "spin" and

"magnetization" are used interchangeably due to the direct relationship between spin angular momentum and magnetization.

To analyze the spin-based thermoelectric transport properties, we synthesized a series of new $Mn_xLi_{1-x}Te$ (x = 0, 0.03, 0.05) samples and characterized them for this study, which reproduced similar transport properties reported as in Ref 37. Details of the synthesis and characterization methods are discussed in the supplementary file. Both undoped and Li-doped AFM MnTe show distinct features in thermoelectric transport properties. The role of Li-doping on carrier transport properties can be understood from the defect equations for Li-doped MnTe:

$$Li(MnTe) \rightarrow Li'_{Mn} + Te^{\times}_{Te} + h^{\cdot}$$

$$Li'_{Mn} + Mn^{\cdot\cdot}_{Mn} \leftrightarrow Li^{\times}_{Mn} + Mn^{\cdot}_{Mn}$$

Here, the notations are the standard for defect equations, i.e., ′ represents the negative charge, × represents the neutral charge, and · represents the positive charge. The subscriptions represent the corresponding sites in the host, and $h$ represents the hole. As seen in the first equation, Li ions in the Mn sites have an effective negative charge inducing a free hole in MnTe to achieve charge neutrality. The subsequent substitution of Li by Mn creates interstitial neutral Li and $Mn^{1+}$.

Figure 2(a)-(b) demonstrates the resistivity and thermopower of $Mn_xLi_{1-x}Te$ samples versus temperature. Both transport properties exhibit distinct features in AFM and PM regimes, which, as will be discussed, are related to spin-based effects. MnTe and $Mn_xLi_{1-x}Te$ samples have p-type electrical properties [37]. The electrical resistivity of both MnTe and $Mn_xLi_{1-x}Te$ is dominated by the spin-disorder scattering, maximized at the transition temperature, $T_N$, above which the electrical resistivities saturate and do not change significantly with temperature [77]. The reduction in resistivity of undoped MnTe above ~600 K is due to the bipolar transport resulting from the thermal activation of electron-hole pairs. As the carrier concentration of doped and undoped MnTe remains the same below and above $T_N$ [37], their electrical conductivity trends follow those of the carrier mobility, mediated by spin-disorder scattering.

Like electrical resistivity, thermopower of doped and undoped MnTe shows spin-effect mediated features, i.e., an advective thermopower contribution with $T^3$ relation from magnon-hole drag effect below $T_N$ [39] and a nearly constant thermopower enhancement in addition to the linearly increasing diffusive thermopower above $T_N$. Li-doped MnTe exhibits a thermopower peak at around 21 K due to the phonon-drag effect [37]. The thermopower enhancement above $T_N$ has been attributed qualitatively to the paramagnon-hole drag [37], although there is yet no theory to formulate it. The decline in the thermopower of the pristine MnTe, again near 600 K, is attributed to the bipolar effect, which was also observed in the electrical resistivity. Li doping in MnTe increases the electrical conductivity and reduces the thermopower while maintaining the spin-mediated thermopower. As a result, $Mn_{0.97}Li_{0.03}Te$ presents nearly twice the larger power factor, $PFT = \alpha^2 T/\rho$, compare to that of MnTe.

According to Figure 2(c), the thermal conductivity ($\kappa$) of both doped and undoped MnTe shows similar trends at low temperatures due to the lattice dominant thermal conduction. With the increase of Li doping, $\kappa$ increases due to the rise of the electronic contribution. Overall, the spin-effects resulted in doubling of $zT$ compared to the case without the spin-effects, where $zT = \alpha^2 T/\rho\kappa$.

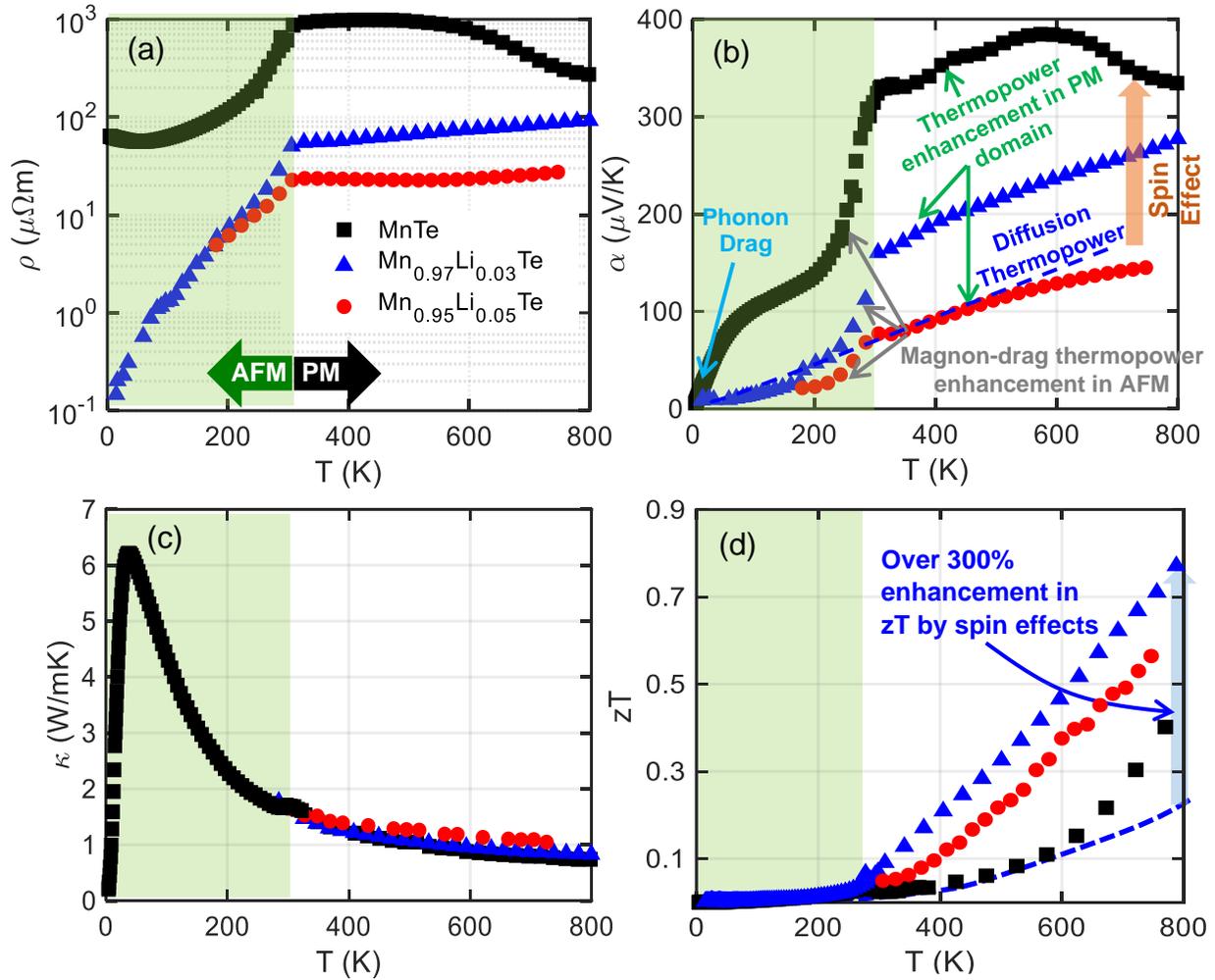

Figure 2: (a) Electrical resistivity, (b) thermopower, (c) Thermal conductivity, and (d) $zT$ of $Mn_xLi_{1-x}Te$ (x=0,0.03,0.05). Spin-based phenomena in AFM and PM provide anomalous excess thermopower contributions in addition to the electronic (or diffusive) thermopower.

As mentioned earlier, the carrier mobility of AFM MnTe is determined by the spin-disorder scattering. Carrier mobility of the MnTe systems is determined from the electrical resistivity and the carrier concentration (from Hall effect measurement) and is illustrated in Figure 5(e). Accordingly, carrier mobility decreases rapidly with the temperature near $T_N$ and remains almost constant above the $T_N$, presumably due to the saturation of the spin-disorder scattering. We will discuss this trend in more detail in the next section.

To investigate the spin-based effects, we will analyze, theoretically and experimentally, the heat capacity, carrier mobility, and relaxation lifetime characteristics of the doped and undoped MnTe samples in the following sections and evaluate the success and limitation of the spin theories in explaining the observed thermoelectric properties.

### Magnon-Carrier Drag Thermopower Below the Magnetic Phase Transition

Understanding various spin-based theories and their application domains is crucial for designing high-performance magnetic thermoelectric materials. However, before going into details on spin theories, it is essential to explain the different spin and carrier relaxation lifetimes used to determine both carrier mobility and drag thermopower. Therefore, four different relaxation processes are introduced.

The first relaxation process is the electron scattering process (relaxation lifetime, $\tau_e$) for the scattering of electrons by everything except magnons. The second mechanism is the magnon scattering with a relaxation lifetime, $\tau_m$, for the scattering of magnons by everything except electrons. The third process is the electron-on-magnon scattering defined by the relaxation lifetime, $\tau_{em}$, which only accounts scattering of electrons by magnons. The last one is magnon-on-electron scattering having a relaxation lifetime, $\tau_{me}$, that considers the scattering of magnons only by electrons. Magnon relaxation lifetime can be calculated from the following expression [78]:

$$\frac{1}{\tau_m} = \frac{1}{\hbar}\sum_i 2J_i S z_i a_i^2 q^2 \qquad (1)$$

$J$ is the exchange interaction energy, $S$ is the spin number, $z$ is the number of nearest-neighbor spins, $a$ is the separation between spins, and $q$ is the magnon wavevector or spin-spin correlation length. Spin-disorder scattering between carriers and magnons causes a momentum transfer from electron to magnon, known as the first-order scattering effect defined by $\tau_{em}$. However, considering the first-order effect alone violates Kelvin's relation, and the contribution from the second-order effect must be included in the formalism. The second-order effect considers that magnons can also return a portion of momentum to the carriers before being randomized [79]. This momentum transfer maintains the equilibrium. From the conservation of linear momentum, and considering only scattering of electrons and magnons, we derive the following expression (see supplementary for derivation):

$$\frac{\tau_{me}}{\tau_{em}} = \frac{k_B T \bar{q}^3}{3\pi^2 p m^* v_m^2} \qquad (2)$$

Eq. (2) defines the relationship between $\tau_{em}$ and $\tau_{me}$. We may assume $\bar{q} \approx \bar{k}$ in this relation. For the case of non-degenerate semiconductors, one can estimate $\bar{k} = k_T = (2m^* kT)^{1/2}/\hbar$, and for the case of degenerate semiconductors, $\bar{k} = k_F = (3\pi^2 p)^{1/3}$, where p is the carrier concentration.

Due to the similar nature of phonons and magnons, the formalism to calculate the magnon-drag thermopower is very similar to that of phonons. The main difference is associated with the nature of the magnons, namely FM magnons versus AFM magnons. Compared to FM magnons, AFM magnon has a linear dispersion typically at long wavelength (like acoustic phonons), degenerate bands, higher magnon velocity, and longer magnon relaxation time. Like carrier mobility, magnon-drag thermopower has contributions from both the first and second-order effects and, therefore, can be expressed as [30]:

$$\alpha_d = \frac{2k_B}{e}\frac{m^* c^2}{k_B T}\frac{\tau_m}{\tau_{em}}\left(1 + \frac{\tau_m}{\tau_{me}}\right)^{-1} \qquad (3)$$

Here, the pre-factor 2 is for the degeneracy of the AFM magnon. The term in the parenthesis of eq. (3) represents the 2$^{nd}$ order drag effect, and the rest is known as first-order magnon drag thermopower ($\alpha_M$), which infers that always $\alpha_d < \alpha_M$. Eq. (3) assumes fixed energy-independent relaxation times. Typically, the first-order drag effect is stronger in AFMs than FMs due to the degenerate magnon modes, higher magnon group velocity, and longer magnon lifetime [39]. On the other hand, FMs have stronger second-order drag thermopower due to the longer magnon-electron relaxation time than that of AFMs [39]. Overall, AFMs exhibit a stronger magnon drag effect due to the dominance of the first-order drag effect over the second-order drag effect [39].

We derive a more accurate approximation following Herring's steps for phonon-drag thermopower [80] as:

$$\alpha_d = \frac{m^*c^2}{k_BT}\frac{\langle\tau_m\rangle}{\langle\tau_{em}\rangle} \times \frac{1}{\frac{m^*c^2\langle\tau_m\rangle}{k_BT\langle\tau_{em}\rangle}\frac{3F(\infty)}{4F(T/T_0)}+1} \times \frac{2k_B}{e} \approx \frac{8k_B}{3e}\frac{F(T/T_0)}{F(\infty)} \quad (4)$$

where $F(x)$ represents the Debye function and $T_0$ is a characteristic temperature which can be determined from: $T_0 = 2\hbar c\, k_F/k_B$, where $k_F$ is the Fermi wavevector. Eq. (4) predicts a maximum limit for the magnon-drag thermopower per magnon mode in the AFM magnon system. For MnTe, $T_0$ is obtained to be about 255 K. Interestingly, magnon-drag thermopower starts showing the enhancement over diffusion thermopower around the same temperature (Figure 2(b)). At this temperature, the average magnon energy is equal to the thermal energy $k_BT_0/2$. Note that for the degenerate semiconductor $\bar{q} \approx k_F$.

The maximum drag thermopower per magnon mode happens when $\frac{m^*c^2\langle\tau_m\rangle}{k_BT\langle\tau_{em}\rangle}\frac{3F(\infty)}{4F(T/T_0')} \gg 1$. According to eq. (4), the theoretical limit for the magnon thermopower of MnTe is about 230 µV/K, considering the second-order drag contribution is insignificant. The inclusion of the second-order effect will reduce magnon-drag thermopower. The experimental magnon-drag thermopower near $T_N$ is unexpectedly close to the theoretical maximum, indicating an over-estimating formalism.

Experimentally, one may estimate the magnon drag thermopower considering $\alpha_d = \alpha - \alpha_e$, where $\alpha_e$ is the diffusion thermopower. For a *p*-type non-degenerate semiconductor, electronic thermopower can be written as [31]:

$$\alpha_e = \frac{k_B}{e}\left(r + \frac{5}{2} + \log(eN_v\mu_t\rho)\right) = \frac{k_B}{e}\log\left(e^{r+5/2}eN_v\mu_t\rho\right) \quad (5)$$

where *r* is the scattering exponent, $N_v$ is the effective hole density of states, $\mu_t$ is the total carrier mobility, and $\rho$ is the resistivity. We plot the experimental $\alpha$ versus $\rho$ for the undoped and two doped MnTe samples (see Figure S3). The $\alpha(\rho)$ plot can be fit with the following function:

$$\alpha = \frac{k_B}{e}\log\left(\frac{\rho}{\rho_0}\right) \quad (6)$$

Here, $\rho_0$ can be determined from the $\alpha(\rho)$ plot, which is ~6.9×10$^{-4}$ Ωcm. Combining eqs. (5) and (6) and inserting $N_v = 2\left(\frac{2\pi m^*k_BT}{\hbar^2}\right)^{3/2}$, one can write another expression for the drag thermopower of a non-degenerate system:

$$\alpha_d = -\frac{k_B}{e}\log\left(e^{r+5/2}eN_v\mu_t\rho_0\right) = -\frac{k_B}{e}\log\left(e^{r+5/2}e\mu_t\rho_0 2\left(\frac{2\pi m^*k_BT}{\hbar^2}\right)^{3/2}\right)$$

$$= -\frac{k_B}{e}\log\left(2e\left(\frac{600\pi m^*k_B}{\hbar^2}\right)^{3/2}e^{r+5/2}\rho_0 x\right) \quad (7)$$

Here, $x = \mu_t \left(\dfrac{T}{300}\right)^{3/2}$ is temperature-weighted mobility. In eq. (7), all the parameters except $x$ are known constants. Figure 3 illustrates the plot of $\alpha_d$ versus $x$ for different values of the hole effective mass at 300 K. It can be seen that the magnon drag thermopower is not a very sensitive function of the effective mass or the carrier mobility for the values of $x < 10$ cm$^2$/Vs. For example, $\alpha_d$ remains around 210-400 µV/K over a large range of the carrier effective mass ($0.3m_e$ - $1.5m_e$) at $x = 2.4$ cm$^2$/Vs (corresponding to the experimental mobility at ~300 K). The sensitivity to the effective mass reduces at smaller $x$ values. Therefore, it can be stated that the magnon drag thermopower near and above $T_N$ is not a strong function of the hole effective mass while the low-temperature thermopower is (see supplementary information). Like the magnon-drag thermopower, magnon lifetime is also a weak function of the hole effective masses for $x < 10$ cm$^2$/Vs at around $T_N$ shown in Figure 3 (see Supplementary information for detailed derivation). However, low-temperature magnon lifetime can change significantly with the hole effective mass.

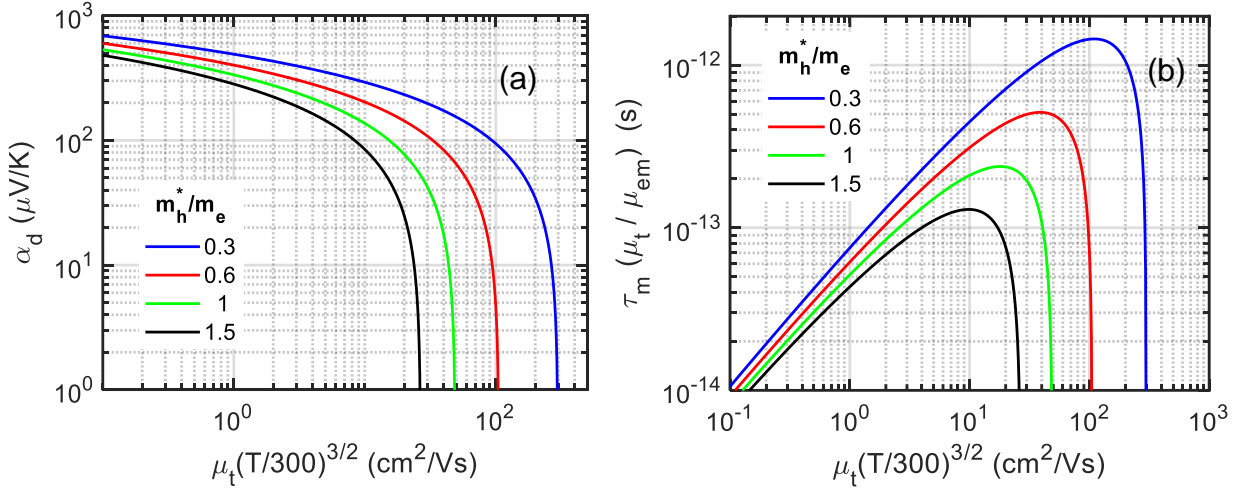

Figure 3: (a) Magnon-drag thermopower and (b) magnon lifetime as a function of temperature-weighted mobility ($x$) for different hole effective masses at T==300 K.

Figure 4 (a) exhibits the experimental thermopower and the empirically extracted diffusion and drag components, while Figure 4 (b) shows the theoretical estimations of the different components. Previously, the thermopower of MnTe was modeled only by the diffusion and magnon-drag thermopowers in the AFM domain [37]. In Figure 4, the theoretical calculations are extended based on the formalisms discussed earlier. Here the $\alpha_{dINS}$ plot is calculated using the magnon lifetime obtained from the inelastic neutron scattering (INS) [37]. The $\alpha_{dtheory}$ plot is calculated using the magnon lifetime obtained from eq. (1). Electronic thermopower, $\alpha_e$, is the same as in Figure 4 (a). Here, we are ignoring the bipolar transport effect at high temperatures. Theoretical maximum drag thermopower, $\alpha_M$, is calculated from eq. (4). According to Figure 4 (b), $\alpha_{dINS}$ and $\alpha_{dtheory}$ are similar, but they both are significantly smaller than the expected drag thermopower, $\alpha_d^{fit}$, in Figure 4 (a). In fact, the drag thermopower, $\alpha_d^{fit}$, is very close to the theoretical maximum thermopower $\alpha_M$. The apparent discrepancy between the theoretical drag thermopower and the experimental value can be due to some of the oversimplifying assumptions, which we will discuss later for the spin-disorder limited carrier mobility theory. Moreover, the magnon-drag described by Eq. (3) is determined from the momentum conservation law between the magnon and carrier systems. Here, the limiting factor can be the consideration of magnon-carrier relaxation processes only; however, there may be multiple magnon-carrier coupling processes involved. Moreover, the theories for determining the various relaxation times involved in the magnon drag thermopower may not be adequately accurate. Another error source can be due to a nontrivial anomalous Hall effect that can affect the carrier relaxation time near $T_N$.

Magnon-drag thermopower can also be determined from the heat capacity data, which results in a better estimation of the drag-thermopower below $T_N$. The thermopower and heat capacity of AFM semiconductors show the same trend of $T^3$ at low temperatures (T<< $T_N$). As will be discussed in the following section, the analysis based on heat capacity can better describe the drag thermopower below the transition temperature.

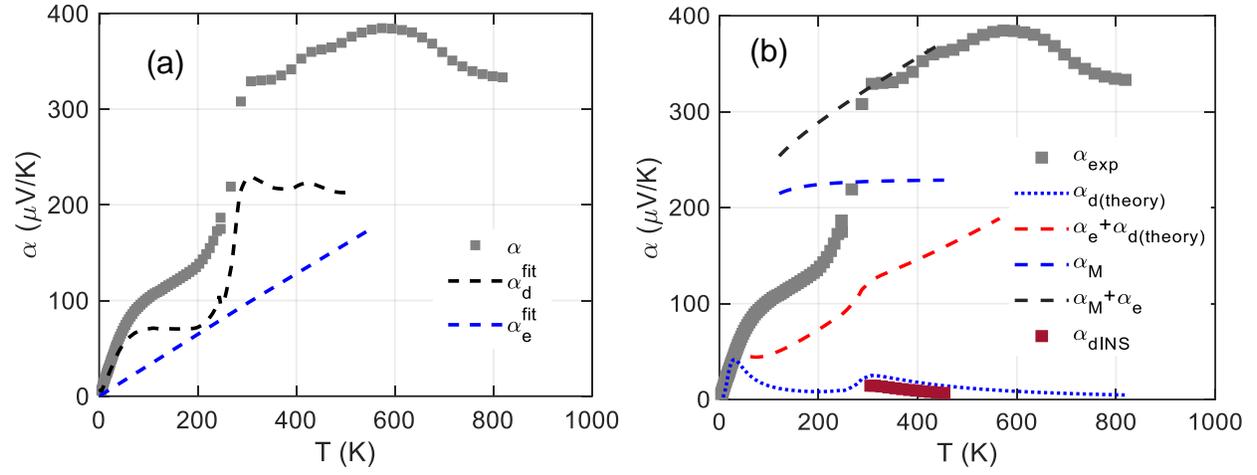

Figure 4: (a) The experimental thermopower (α) of MnTe and the fitting components: diffusion ($\alpha_e^{fit}$) and drag ($\alpha_d^{fit}$). (b) Experimental thermopower, $\alpha_{exp}$, along with theoretically estimated drag contributions, $\alpha_{d(theory)}$ and $\alpha_e + \alpha_{d(theory)}$, theoretical maximum drag thermopower, $\alpha_M$ and $\alpha_M + \alpha_e$, and the drag thermopower calculated from the magnon lifetime obtained from INS measurement, $\alpha_{d(INS)}$.

## Discrepancy of the Magnon/Paramagnon Lifetimes from INS and Transport Data

As discussed, magnon/paramagnon lifetimes play a crucial role in determining the spin-based thermoelectric transport properties near and above the transition temperature. Earlier sections discussed the theoretical model to determine these lifetimes, which can be different from the practical values due to the limitations highlighted in the previous section. Therefore, in this section, we discuss the empirical methods to determine the magnon and paramagnon lifetimes.

Inelastic neutron scattering (INS) is a direct method for estimating the spin relaxation time. With this technique, the neutron intensity of inelastic scattering by magnons or magnetic fluctuations is measured as a function of energy ($E$) and momentum transfer ($Q$), where $|Q| = 4\pi \sin\theta/\lambda$. Here, $2\theta$ is the scattering angle, and $\lambda$ is the neutron wavelength. Generally, for single crystals, the lifetimes for both magnons and phonons with specific wavevectors can be calculated from the intensity of the INS scattering function, $S(Q,E)$. Both inelastic and elastic features can be present in the $S(Q,E)$ plots. The lifetimes are calculated from the full-width at half maximum (FWHM) of the Lorentzian-fitted inelastic features using the Heisenberg energy-time uncertainty principle: $\Delta E \cdot \Delta \tau \approx \hbar$. Magnon lifetimes for a polycrystalline sample cannot be determined from INS due to the random orientation. In the paramagnetic domain, a broad inelastic feature centered on $E = 0$, called quasielastic scattering, can be used for determining the paramagnon lifetimes, as magnons cease to exist and are replaced by liquid-like magnetic fluctuations above the transition temperature. To be precise, this is the relaxation rate in the spin-spin pair-correlation that is determined. This approach is also applicable for polycrystalline material, though some information on the directionality of spin-fluctuations is lost. Similarly, the spin-spin correlation *length* can be roughly estimated in the orientational average for a polycrystal from the FWHM of the broad feature that replaces the magnetic Bragg peak, calculated from an intensity vs. momentum plot, $S(Q)$, using the Heisenberg uncertainty principle: $\Delta x \cdot \Delta p \approx \hbar$.

For undoped MnTe, INS was performed on a ~10 g pressed pellet at the Wide Angular Range Chopper Spectrometer, ARCS, of the Spallation Neutron Source at Oak Ridge National Laboratory using neutrons with 60 meV incident energy. Data were analyzed to estimate the magnon/paramagnon lifetimes. The INS spectra at different temperatures above and below $T_N$ are illustrated in Figure 5(a). In Ref. 37, the paramagnon lifetimes were reported only for Li-doped MnTe. Here, we measured the lifetimes using INS for the undoped MnTe as well, as shown in Figure 5.

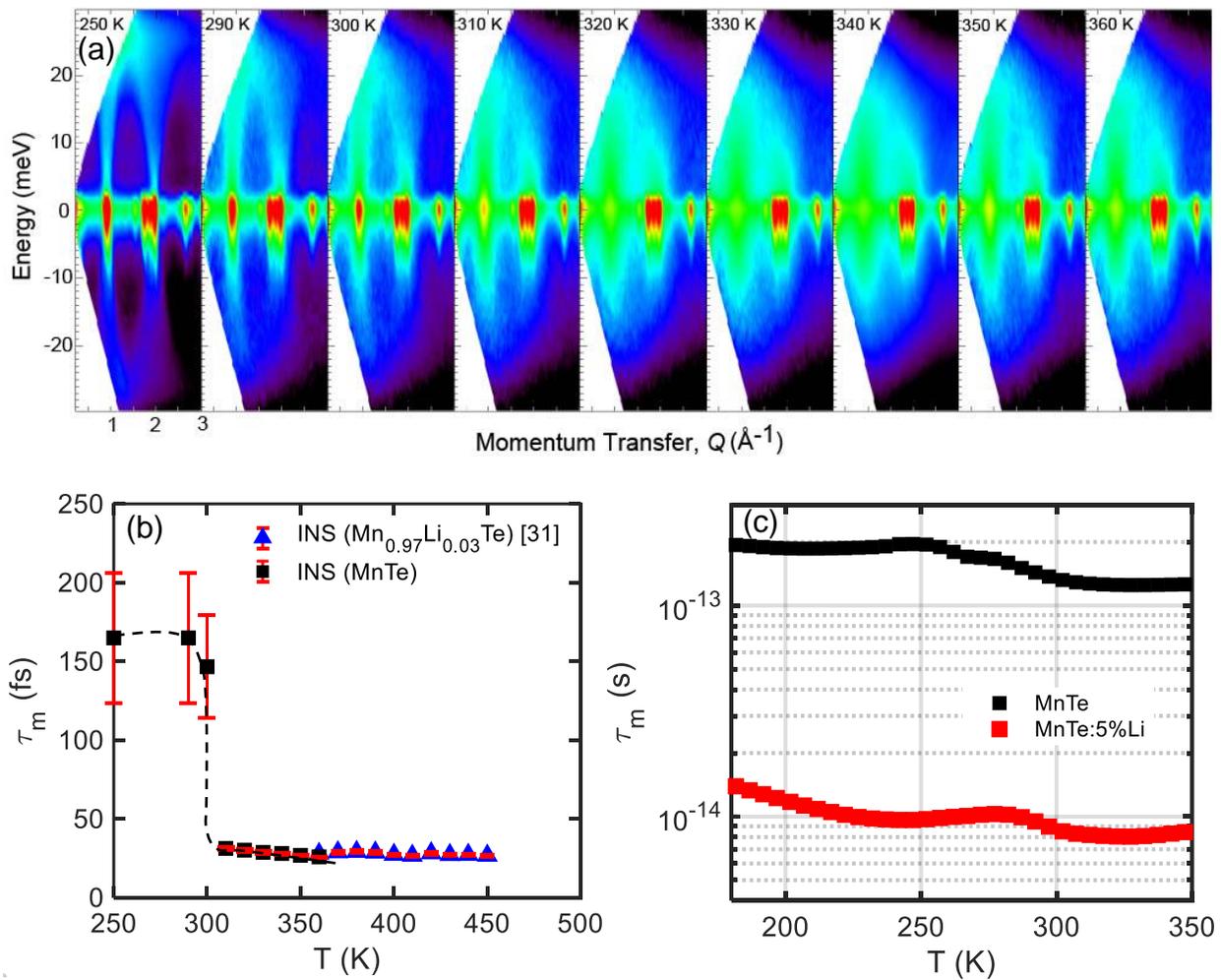

Figure 5: (a) Illustration of neutron spectra (*S(Q,E)*) from INS measurements at different temperatures, (b) calculated magnon/paramagnon lifetime from the transport properties, and (c) the estimated magnon/paramagnon lifetimes of MnTe and 3%Li-doped MnTe [ref. 37] from neutron spectra.

The undoped MnTe shows distinct magnetic Bragg peaks at around 0.92 Å$^{-1}$ and 1.95 Å$^{-1}$ below $T_N$, along with the magnon bands extended up to around 30 meV, which is in agreement with the previous neutron studies [37,74]. Above $T_N$, the distinct magnon bands disappear, and a broad feature associated with paramagnon exists at 0.92 Å$^{-1}$. The feature representing the paramagnon scattering remains unchanged in intensity and energy distribution at all temperatures in the PM domain. The paramagnon lifetimes are estimated from the Lorentzian-fitted quasielastic features of *S(E)* obtained from a slice at 0.92 Å$^{-1}$, see Ref 37. In comparison, another slice at 1.5 Å$^{-1}$ is considered to estimate the magnon lifetime near the van Hove singularity from the Lorentzian-fitted inelastic features in *S(E)* at ~25 meV. Note that the actual magnon lifetime can be higher than the estimated lifetimes below the transition temperature, while paramagnon lifetime estimation can be closer to the practical values. The obtained magnon/paramagnon lifetimes are shown in Figure 5(b). It can be seen that the paramagnon lifetime is approximately constant above $T_N$ and is about ~30 fs.

Next, we derive the lifetime expression from the drag thermopower and carrier mobility. We assume that electrons are dominantly scattered by magnons, ignoring other scattering mechanisms, and rearrange the drag mobility expression from Ref. 78 as:

$$\tau_{em}\left(1+\frac{\tau_m}{\tau_{me}}\right) = \frac{m^*\mu_d}{e} \qquad (8)$$

Inserting eq. (8) into eq. (3), we have:

$$\alpha_d = \frac{2k_B}{e}\frac{em^*c^2}{k_BT}\frac{\tau_m}{m^*\mu_d} = \frac{2c^2\tau_m}{T\mu_d} \qquad (9)$$

By rearranging eq. (9), the expression for magnon lifetime can be obtained as:

$$\tau_m = \frac{T\mu_d\alpha_d}{2c^2} \qquad (10)$$

Eq. (10) can be used to estimate the magnon relaxation time from the experimental magnon carrier drag thermopower $\alpha_d$ and the Hall mobility $\mu_d$. One may derive a more accurate relation by taking the energy-dependent relaxation times and solving the coupled Boltzmann transport equations for electrons and magnons similar to Herring's treatment for phonon drag effect [81]. Considering the magnon velocity as 14000 m/s [78], we can estimate the magnon lifetime for undoped and 5%Li-doped MnTe from Eq. (10), as illustrated in Figure 5(c).

The magnon lifetime derived from the transport properties of undoped MnTe is in the same range as the one estimated from the INS data. The paramagnon lifetimes of undoped and doped MnTe obtained from INS are within the same range but orders of magnitude smaller than the lifetime calculated from the transport properties of undoped MnTe.

In order to explain the experimental paramagnon drag thermopower data, the parmagonon lifetime must be in the range of 120 fs. However, the lifetime estimated by INS is smaller in the range of 30 fs. The difference can be associated with assuming a constant magnon velocity when evaluating eq. (10). As such, temperature-dependent magnon/paramagnon velocities below and above $T_N$ have been used to reconcile with neutron scattering experiments (refer to Eq. 14 of the following section) [82].

**Magnon/Paramagnon-Drag Thermopower from Magnon Heat Capacity**

Heat capacity ($C_p$) can reveal different phase-transitions and entropy carriers like phonons, charge carriers, Schottky, hyperfine, magnon, and spin-transition. As such, the temperature-dependent features associated with those contributors can be tracked in the heat capacity trend [73]. Both the heat capacity and thermopower of a system are thermodynamically related. The magnon-drag thermopower can be approximately calculated from the magnon heat capacity using the following relation [37]:

$$\alpha_d = \frac{2}{3} \frac{C_m}{ne} \frac{\tau_m}{\tau_m + \tau_{me}} \qquad (11)$$

The term containing the relaxation times $\tau_{me}$, $\tau_{em}$, and $\tau_m$ take into account the fraction of the momentum that transfers from the magnons to charge carriers. In general, this ratio is a function of various parameters like magnetic ordering, degenerate or non-degenerate semiconducting nature, defects, temperature, etc. The limitations of models for calculating the lifetimes can introduce a discrepancy between the experimental and theoretical values. For example, previous literature demonstrated success in explaining only the experimental drag thermopower trends using the heat capacity data below transition temperature; however, it failed in explaining the numerical values [37], and an arbitrary coefficient was used to scale and fit the data [37]. However, with no theoretical basis for the arbitrary coefficient, the analysis can mislead physical explanations.

First, it is essential to model the magnonic heat capacity correctly to evaluate the drag thermopower. The experimental and theoretical heat capacity of MnTe is illustrated in Figure 6 with the contributions from different entropy carriers, namely phonons ($C_v$), dilation ($C_d$), Schottky ($C_{Sc}$), and magnons ($C_m$). It should be noted that Li doped MnTe exhibits similar heat capacity trends. The magnon heat capacity contribution in MnTe has also been reported earlier [37,73,83]. By considering the various parameters associated with the non-magnetic heat capacity contributions from Ref. 84, one can determine the magnonic contribution to heat capacity from the experimental values. According to Figure 6, magnon heat capacity contribution maximizes at Néel temperature. Below $T_N$, both magnon heat capacity and magnon-drag thermopower show a similar $T^3$ trend [37]. According to the spin-wave theory [85], spin-waves are well defined up to the transition temperature due to long-range ordering that bestows rigidity. Here, it is essential to note that the spin-wave contribution to heat capacity calculated from linear spin-wave theory shows a $T^3$ trend at lower temperatures than $T_N$ and saturates near $T_N$, like phonon heat capacity [86] (see supplementary). Therefore, the linear spin-wave theory cannot explain the λ-anomaly at $T_N$. Long-range magnetic ordering breaks into mid to short-range ordering above the transition temperature and introduces λ-magnetic phase transition [86]. Breaking of the magnetic order can also contribute to the heat capacity along with spin-wave due to the entropy change [87]. Spin-wave heat capacity generally has an asymmetric trend below and above the transition temperature, while magnetic ordering-disordering can have a symmetric trend [86]. Typically, magnetic semiconductors exhibit λ-anomaly in heat capacity, attributed to both the spin-wave and magnetic ordering-disordering entropy [86].

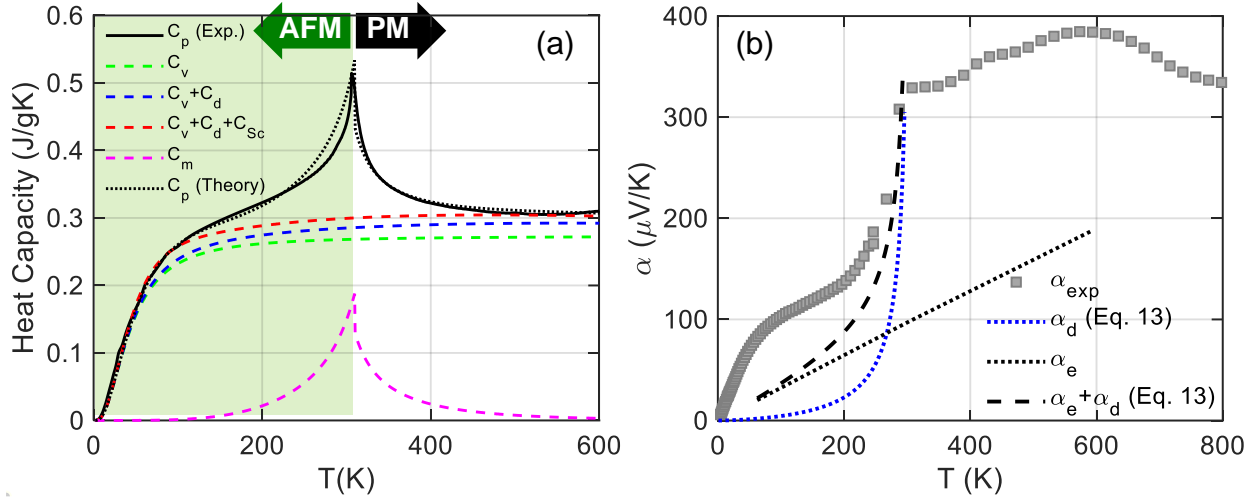

Figure 6: (a) Specific heat capacity ($C_p$) of MnTe along with its contributing components: the phonon($C_v$), dilation ($C_d$), Schottky ($C_{Sc}$), $C_v+C_d+C_{Sc}$, and magnon ($C_m$). Experimental $C_p$ (solid black) is compared with theoretically calculated $C_p$ (dotted black). (b) Experimental thermopower and the calculated drag thermopower following Siguhara's formalism [82].

The well-known AFM magnon heat capacity based on spin-wave theory is given below [85]:

$$C_{mag} = CNk_B \left(\frac{k_B T}{2JS\sqrt{2z}}\right)^3 \qquad (12)$$

The constant $C$ is a lattice structure-dependent parameter. $N$ is the total number of magnetic ions, $k_B$ is the Boltzmann constant, $J$ is the exchange energy, $z$ is the number of nearest-neighbors, and $S$ is the ground state spin number. Equ (12) is valid only at low temperatures and cannot explain the AFM magnon heat capacity near $T_N$. On the other hand, the λ contribution to the heat capacity due to the order-disorder transition is modeled by the relations based on several fitting parameters [87]. Therefore, it is imperative to accurately calculate the magnon heat capacity to determine the magnetic contribution to the heat capacity accurately. The low-temperature magnon heat capacity can be directly estimated based on the linear spin-wave theory from the magnon density of states. However, the linear spin-wave theory breaks down near the Néel temperature; as such, special attention must be taken to use an accurate model for the magnon heat capacity near the transition temperature.

Considering the two magnetic heat capacity contributions, spin-wave, and magnetic order-disorder transition, it is also essential to know which heat capacity should be used to determine the magnon drag thermopower based on Eq. (11). Due to the λ-anomaly observed in MnTe, it is expected that the spin-wave heat capacity contribution in MnTe is smaller than the $C_m$ shown in Figure 6. In previous literature [37], the magnon-drag thermopower was estimated using eq. (11) by considering total magnon heat contribution ($C_m$) and the $\tau_m/(\tau_m + \tau_{me})$ term as a fitting parameter. The latter term was assumed to be 1/100, which does not agree with the lifetime estimates of Figure 9. Therefore, an accurate estimation of the spin-wave heat capacity contribution ($C_{mag}$) may better estimate the magnon-drag thermopower.

K. Sugihara provided another formalism based on magnon heat capacity to estimate the magnon drag thermopower. In this formalism, a magnon mode-dependent heat capacity is multiplied by a momentum transfer ratio and summed over all magnon modes [82,84]. The magnon-drag thermopower in this formalism is expressed as [82]:

$$\alpha_d = -\sum_q \frac{R(q)c_m(q)}{3e} \quad (13)$$

Where $e$ is the electron's charge, $q$ is the magnon wavevector, and $C_m$ is the magnon's specific heat. $R$ is the momentum transfer ratio between the magnon and electron systems, which can be determined from the total magnon relaxation time over the relaxation time of magnons due to *s-d* interaction [82]. The drag thermopower from eq. (13) is also limited to the Néel temperature and shows a divergence at $T_N$. The thermopower obtained from eq. (13) is illustrated in Figure 6(d) for comparison.

Both thermopower models can explain the experimental data by using some fitting parameters. The model described in eq. (11) uses an arbitrary value for $\tau_m/\tau_{me}$ to explain the thermopower data. Sugihara's model calculates the lifetime without fitting parameters; however, the model assumes a temperature-dependent magnon velocity $v_s(T)$ to reconcile with neutron scattering experiments:

$$v_s(T) = v_s(T_N)\left[1 + \alpha_i(1 - T/T_N)^{\delta_i}\right]$$
$$i = \begin{cases} 1: T \lesssim T_N \\ 2: T \gtrsim T_N \end{cases} \quad (14)$$

Here $\alpha_i$ and $\delta_i$ are fitting parameters.

While both models can explain the thermopower trend below $T_N$, using some fitting parameters, the thermopower above $T_N$ cannot be explained with either model. Therefore, the thermopower trend in the PM domain requires further investigation.

**Limitations of the Paramagnon Drag Thermopower Formalism**

As shown in Figure 2(b), the thermopowers of MnTe and Li-doped MnTe samples keep increasing without a decline above the transition temperature. This anomalous trend of the thermopower above the magnetic transition temperature was hypothetically attributed to the paramagnon electron drag [37]. Inelastic neutron scattering (INS) measurements have been used to find the magnon lifetime above the Néel temperature for both undoped and doped MnTe [37]. From the inelastic neutron scattering (INS), paramagnon lifetime and energy spreading have been measured at different temperatures, and the estimated paramagnon lifetime was found to be ~30 fs. The corresponding paramagnon correlation length was found to be ~2-2.5 nm, which is higher than the free-carrier effective Bohr radius (~0.5 nm) and de Broglie wavelength ~0.6-1 nm [37]. It was argued that since the spatial extent of the thermal fluctuation of magnetization is larger than the effective Bohr radius of the electron, paramagnons appear as magnons to electrons; hence, paramagnons can induce a similar drag effect to that of magnons.

The problem with this argument arises by noting the magnon lifetime's discontinuity, as seen in Figure 5(c). There is an apparent drop in the paramagnon lifetime compared to the magnons at around $T_N$; however, the drag thermopower does not show any drop at the phase transition temperature.

Using the experimental lifetime of the MnTe system from INS in the paramagnetic domain, paramagnon-drag thermopower is calculated from eq. (3) where $\tau_{em}$ is taken from the Hall mobility data, and $\tau_{me}$ is obtained from eq. (2). As shown in Figure 4, the calculated paramagnon-drag thermopower is significantly smaller than the experimental value. The calculated paramagnetic thermopower is dominated by the diffusion thermopower (Figure 4(b)), while the experimental decomposition (Figure 4(a)) shows a significant component in addition to the diffusion thermopower in the PM domain. The discrepancy observed in the paramagnetic domain suggests at least two possible explanations: (i) the drag thermopower requires a new formalism to explain the experimental data, or (ii) the MnTe system has different or multiple spin-based mechanisms that may or may not include the paramagnon-drag effect.

As discussed, above the Néel temperature, magnons are broken into paramagnons with shorter lifetimes, and long-range magnetic ordering is broken into short- to mid-range magnetic ordering. Both of them can create excess entropy in the disordered domain. We can calculate the magnetic entropy related to the spin-wave by using the relation:

$$S_M = \int \frac{C_{mag}}{T} dT \qquad (15)$$

Here, $C_{mag}$ is the spin-wave (magnon) heat capacity, and $S_M$ is the associated magnetic entropy. In the high-temperature limit, we can relate the magnetic entropy to thermopower by $\alpha_{ME} = S_M/ne$, where $n$ is the concentration of the charge carriers. This estimation of magnetic thermopower contribution in the PM domain needs an accurate estimation of $C_{mag}$.

In summary, per the observations from the experimental and theoretical data in AFM and PM domains, one can make the following conclusions. If the paramagnon electron drag thermopower follows a similar formalism as that of magnons, the thermopower must show a decline right above $T_N$ as one would expect an abrupt reduction of $\tau_m$ when the material transitions from the AFM to PM phase [88]. However, the experimental data exhibits a constant contribution (with no decline) to the thermopower in the PM domain in addition to the diffusion thermopower. The magnetic entropy shows a similar trend. These conclusions suggest that either the paramagnon drag thermopower needs more accurate governing formalism, or the paramagnetic thermopower (excluding the electronic thermopower) must have contributions from multiple spin-based transport processes, such as a combination of paramagnon-drag and magnetic entropy.

In the following section, we will explore the spin entropy prospect originated from the degeneracy of two different spin centers to explain the paramagnetic thermopower. Here, it is essential to differentiate the terms spin entropy and magnetic entropy from spin-wave. A straightforward difference is that spin entropy is caused by the presence of two different magnetic centers in the host, but spin-wave entropy is caused by the randomness of the spin-wave.

**Spin Entropy to Explain Excess Thermopower**

According to the thermodynamic relation, thermopower can be considered as the amount of entropy carried by a unit charge carrier in the direction of the charge flow [38]. The entropy gradient creates a flow of the electronic fluid in the solid making a net electric field by repositioning the Fermi energy to balance the net current flow. The entropy gradient in the system can originate from various sources, such as the thermal gradient or the spin/orbital degeneracy of various magnetic centers in the system. The entropy from spin and orbital degeneracy can cause the relocation of electrons from a high entropy state to a low entropy one. Spin carried entropy is often insignificant in many compounds except in materials with strong electron-electron interactions such as transition-metal systems [32-34]. In a transition-metal system, 3d electrons can have both spin and orbital degeneracy originated from the degeneracy of the electronic spin states of the magnetic ions, namely, low, intermediate, or high spin states. Such electronic configurations in 3d orbitals are primarily due to the competition between the crystalline field and the Hund's rule coupling. Based on the Heikes formula [89] at the high-temperature limit, thermopower or Seebeck coefficient due to the crystal field-driven spin entropy can be expressed as [40]:

$$S = \frac{\mu}{eT} = -\frac{\sigma}{e} = -\frac{k_B}{e}\ln(g_s g_c) \quad (16)$$

where µ is the chemical potential and σ is the entropy per electron, which equals energy per unit temperature coming from the spin entropy, $g_s$ is the spin degeneracy, and $g_c$ is the configurational degeneracy. $k_B/e \approx 86.25$ µV/K can be considered as the natural unit of the thermopower. The necessary condition for electron hopping in a system with spin entropy is that the change in the total spin number of the system should be zero.

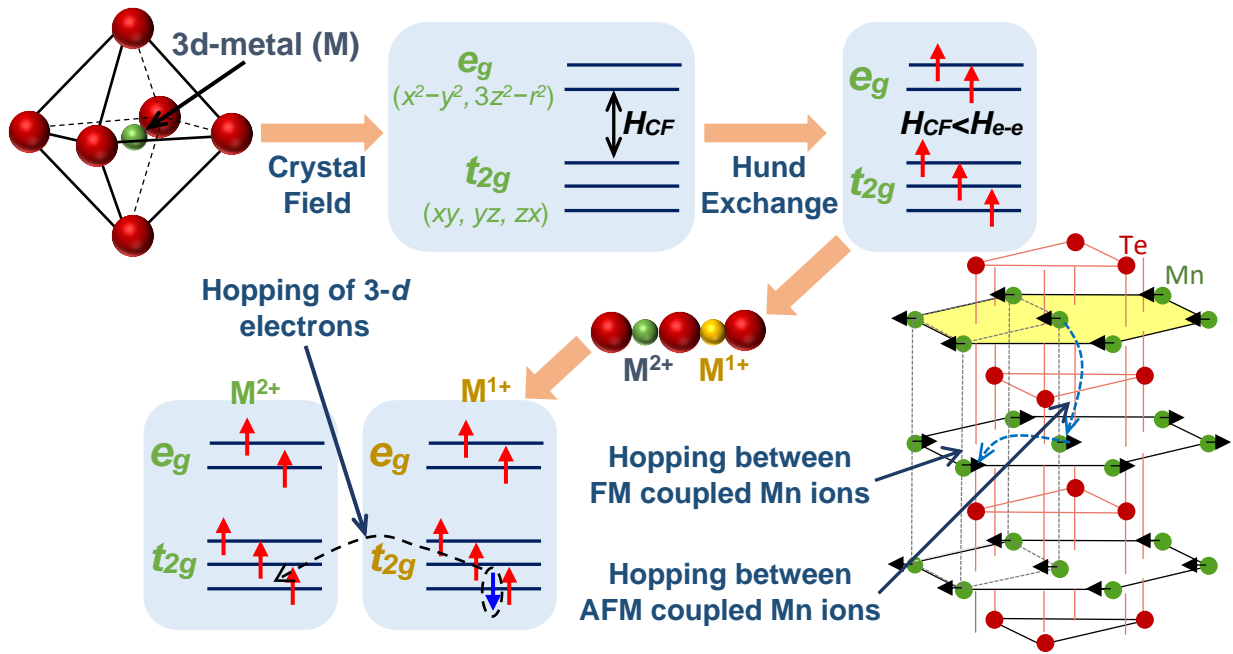

Figure 7: Electron hopping between $Mn^{2+}$ and $Mn^{1+}$ ions of the strongly correlated MnTe system due to the spin entropy gradient coming from different spin and configurational degeneracies. $H_{CF}$ and $H_{e-e}$ are associated with the crystal field and Hund's rule coupling energies.

The spin entropy was found to explain the anomalously high thermopower in metallic oxide sodium cobaltate ($Na_xCo_2O_4$)[40]. In the MnTe system, we show that the spin degrees with strong p-d interaction may also be responsible for the large thermopower contribution. If we assume a hole transport similar to metallic $Na_xCo_2O_4$, MnTe can exhibit the spin entropy thermopower at the presence of $Mn^{2+}$ and $Mn^{1+}$ ions. The existence of $Mn^{1+}$ ions in MnTe has already been confirmed by the XPS studies, which can be due to the broken Mn-Te bonds [90]. The concentration of $Mn^{1+}$ and $Mn^{2+}$ ions can be determined from high-temperature magnetometries like magnetic susceptibility measurement. As we will show, the presence of even a small amount of $Mn^{1+}$ ions (a few percentages) in $Mn^{2+}$ host can provide a significant impact. When electrons from the $^6A_1$ ground state of $Mn^{2+}$ ion ($3d^5$) move to $Mn^{1+}$ ions ($3d^6$) with $^7S_3$ state, the total change in the spin number remains zero; hence, it satisfies the condition for spin entropy-initiated electron hopping. It is expected that due to the strong correlation, holes can hop between the empty states (Figure 7) and contribute to the thermopower. Here, it is essential to identify the physical sources of different Mn ions. The hybridization of the *p(Te)-d(Mn)* and *s(Te)-d(Mn)* orbitals can play a crucial role in the interaction of the itinerate electrons and the 3d electrons. A recent article also discusses the spin entropy thermopower in MnTe, relating it to the delocalization of d-electrons due to the band-hybridization [91]. As mentioned earlier, both conduction and valence bands of MnTe are hybridized by Mn 3d bands and the Te 5p and 5s bands [92,93,94]. Doping and defects can both induce different types of magnetic ions in the system. Any charge-transfer reaction between the ions can introduce new electronic configurations into the system [73]. The intrinsic *p*-type conductivity of MnTe can be explained by the Mn vacancies [31], which can introduce various magnetic centers into the system.

According to eq. (16), spin entropy thermopower is a strong function of the configurational and spin degeneracy of the system. The configurational degeneracy is due to all the atoms involved in hopping, which are $Mn^{2+}$ and $Mn^{1+}$ in the case of MnTe. The spin entropy is due to both the spin degeneracy and orbital configuration degeneracy [32,34]. Therefore, to calculate the spin entropy thermopower, one must determine both. For instance, considering spin degeneracy for high-spin (HS) $Mn^{2+}$, $g_2 = 6$, and for HS $Mn^{1+}$, $g_1 = 15$ (see Figure 7). In MnTe, electronic states are singly occupied due to the large on-site electron-electron repulsion energy (*U*). If *t* is the hopping integral, the condition of $k_B T \gg t$ meets the the high-temperature limit of the Hubbard model [32,34]. Therefore, the total configurations number (*g*) is limited by the condition of $U \gg k_B T \gg t$. This condition also means that the carriers are free to move between the sites – not bound to hoping energy *t*. From the total number of configurations under the given condition, the spin thermopower at the high-temperature limit can be written as [32,34]:

$$\alpha = -\frac{k_B}{e}\frac{\partial \ln g}{\partial N} \qquad (17)$$

Here, *N* is the number of electrons. Considering the existence of $Mn^{2+}$ and $Mn^{1+}$ in the MnTe system, the total number of configuration (*g*) can be obtained as [32,34]:

$$g = g_s g_c = g_2^{N_T-M} g_1^M \frac{N_T!}{M!(N_T! - M!)} \qquad (18)$$

Where, $N_T$ is the total number of Mn ions in MnTe, and *M* is the number of $Mn^{1+}$. Inserting eq. (18) in eq. (17), and using Stirling's approximation, one can arrive at the following simplified formula for the spin entropy thermopower:

$$\alpha = -\frac{k_B}{e} \ln\left(\frac{g_2}{g_1}\frac{x}{1-x}\right) \qquad (19)$$

where $x = \frac{Mn^{1+}}{Mn}$. If we assume that the concentration of $Mn^{1+}$ ions is about 2% of that of Mn ions, i.e., $x = 0.02$, the spin entropy thermopower is around 410 µV/K. This value is larger than the maximum excess thermopower measured in MnTe. Considering that MnTe is not a hopping system, this approach may overestimate the spin entropy thermopower. Nevertheless, it can offer an alternate explanation for the anomalous trend of the thermopower in the PM domain.

According to the previous literature [40,95,96,97], a strong magnetic field-dependent thermopower is a signature of spin entropy thermopower. For example, the magnetothermopower of $Na_xCo_2O_4$ shows a substantial reduction with increasing the field [40]. To check the effect of the magnetic field on the thermopower of MnTe, we measure the magnetothermopower along with field-dependent magnetic moment (*M-H*) and heat capacity (shown in supplementary). The results show that the external field almost has no impact on the thermopower up to 12 T. The impact of the external field on spin entropy contribution depends on the material. If the external field is strong enough to force the spin alignments in the direction of the field, the spin entropy can be affected by the external field. A similar discussion is made in [40] when comparing the in-plane and c-axis field-oriented data. However, Li-doped MnTe shows almost no variation in thermopower or heat capacity with the field, indicating that the 12 T is insufficient to change the spin alignment and affect the spin entropy contribution. The MH plot also shows an almost featureless linear trace with minor changes in the slope.

MnTe and $Na_xCo_2O_4$ [40] are two different systems in terms of magnetic properties. Sodium-cobaltate is a paramagnetic metal oxide with a large magnetic susceptibility, about two orders larger than MnTe (and typical metals). That explains the field-dependent spin properties even at low fields. For the case of MnTe, as shown in the MH plot, the magnetization of the sample is mostly compensated, and only a small fraction of the Bohr magneton is detected even above the spin-flop transition. The net magnetization is only ~0.04µB per Mn at the 12 T field. This corresponds to less than a 1° canting angle. In this respect, MnTe is an entirely different case than sodium cobaltate. Equ. (2) in ref [40] described the field-dependent thermopower based on the assumption of non-interacting residual free spins. It works for solidum cobaltate but does not apply to MnTe. A relevant observation is the paramagnetic curie temperature ($\theta$) of sodium cobaltate (~55K), which is an order of magnitude smaller than MnTe (~575K). As the antiferromagnetic exchange energy is $J_{AF} \approx \theta$ [40], that also explains a much larger field is needed to change the spin alignment and eliminate the spin entropy in MnTe.

**Spin-Disorder Scattering to Explain Carrier Mobility**

Different spin disorder scattering models by C. Haas [77], G. Zanmarchiand C. Hass [78], and C. Herring [80] tried to explain the carrier mobility of undoped and doped MnTe systems with limited success. All models considered that the itinerate carriers are scattered by the random motion of the spin of lattice ions. This scattering is primarily dominated by the *s-d* or *p-d* exchange interaction. It should be noted that not all magnons can scatter charge carriers. The problem is similar to the case of electron scattering by acoustic phonons [98], where there is a limit to the wavevector of the magnons that can scatter an electron with wavevector $k$, i.e., $q \leq q_m = 2k + \frac{m^* v_m}{\hbar}$. This relation indicates that the energy of an electron can change after scattering by magnons at most by $\hbar v_m q_m = 2\hbar v_m k + m^* v_m^2$. Comparing this energy with $kT$, one can find the temperature $T_i$ below which the effect of inelastic scattering may become significant (similar to the case of Debye temperature) [99].

For the doped sample $Mn_{0.97}Li_{0.03}Te$, assuming $p \cong 4.5 \times 10^{20}$ cm$^{-3}$, one has $T_i \cong 507$ K. Therefore, for the case of doped MnTe, the effect of inelasticity can be significant for a broad range of temperatures that go above $T_N$. This infers that the elastic scattering-based models are not valid in the broad range of temperature, indicating a major shortcoming in the models. Nevertheless, it has often been ignored in the previous literature [77,78].

Carrier mobility, due to the drag effect, including both the 1$^{st}$ and 2$^{nd}$ order effect, can be written as [78]:

$$\mu_d = \frac{e\tau_{em}}{m^*}\left(1 + \frac{\tau_m}{\tau_{me}}\right) \quad (20)$$

The term in the parenthesis represents the 2nd order effect. This term was ignored in the calculation of spin-disorder mediated carrier mobility in previous literature [77,80]. However, it can significantly impact the carrier mobility when the carrier diffusion thermopower becomes comparable to the magnon-drag thermopower, like in the MnTe system.

In a magnon system with itinerant electrons, spin-disorder scattering can be associated with spin-flip or spin non-flip scattering [39,77]. Both energy-dependent scattering mechanisms depend on magnetic susceptibility and magnon band structure [39,77]. Spin-flip scattering is a two-magnon process that occurs only in the AFM magnon system due to the degenerate magnon band [39]. On the other hand, spin non-flip scattering is a one-magnon scattering process that can exist in both AFM and FM materials [39]. In the spin-disorder scattering theory, relaxation time can be calculated for both spin-flip ($\tau_{\uparrow\downarrow}$) and spin non-flip ($\tau_{\uparrow\uparrow}$) scattering processes. The energy dependent lifetimes can be determined from the following relations [77]:

$$\tau_{\uparrow\downarrow}(E) = \frac{\hbar(2Ng\mu_B)^2}{2\pi J^2 k_B T} \left( 2\chi^\perp \sum_{k'} \delta(E_k^\pm - E_k^\mp) \right)^{-1} \quad (21)$$

$$\tau_{\uparrow\uparrow}(E) = \frac{\hbar(2Ng\mu_B)^2}{2\pi J^2 k_B T} \left( \chi^\parallel \sum_{k'} \delta(E_k^\pm - E_k^\pm) \right)^{-1} \quad (22)$$

Here, $N$ is the number of the magnetic ions per unit volume, $g$ is the g-factor, $\mu_B$ is the Bohr magnetron, and $\chi$ is the magnetic susceptibility. For calculating the spin disorder relaxation time, magnetic susceptibility is considered as $(\chi^\parallel + 2\chi^\perp)$. $\chi^\parallel$ and $\chi^\perp$ are taken from experimental data, which are given in the supplementary information.

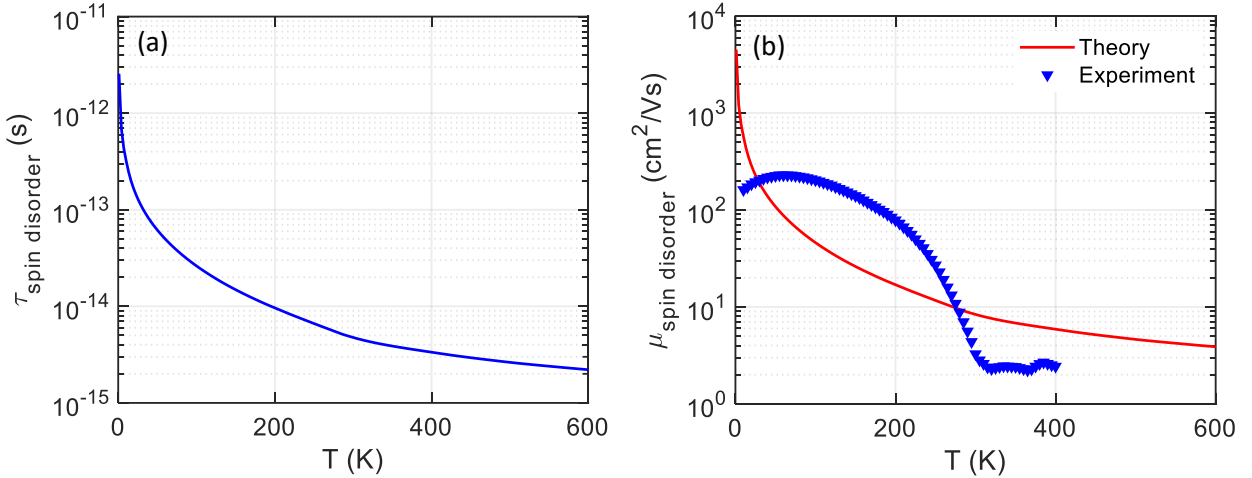

Figure 8: (a) Calculated temperature-dependent spin-disorder scattering lifetime for MnTe assuming a hole concentration of $10^{19}$ cm$^{-3}$. (b) Carrier mobility of MnTe from experiment and spin-disorder scattering theory according to [77].

Using the above relations, one can calculate spin-flip, spin non-flip, and spin-disorder scattering lifetimes. Energy-dependent spin-flip and spin non-flip scattering lifetimes are shown in supplementary, while spin-disorder scattering lifetime is illustrated in Figure 8 (a). The spin-disorder scattering relaxation time ($\tau_{\text{spin disorder}}$) is in the order of several fs between 200K-1000K. This is some orders of magnitude larger than typical relaxation times, such as those due to scattering by acoustic phonons and ionized impurities. Therefore, one can assume that the total hole relaxation time is approximately the same as $\tau_{\text{spin disorder}}$ and can calculate the carrier mobility for degenerate AFM MnTe. Figure 8(b) compares the theoretical carrier mobility with the experimental values (see supplementary for the analysis). As seen in the figure, spin-disorder theory is unable to capture the trend of carrier mobility. The calculated mobility is different than the observed value for most of the temperature range, which indicates the limitation of the current theory. As mentioned earlier, the carrier mobility of the MnTe system is influenced by both first-order and second-order effects. The second-order effect enhances the carrier mobility; therefore, the difference can be partially explained by the absence of the second-order contribution to this theory.

We adopt an alternative route to find out the impact of the 2$^{nd}$ order effect on carrier mobility. From eq. (2), the $\tau_{me}$ can be written as:

$$\tau_{me} = \frac{\tau_{em} k_B T k_{avg}^3}{3\pi^2 p m^* v_m^2} \qquad (23)$$

Here, $k_{avg} \approx q$ near the Fermi energy [78]. By inserting eq. (23) into eq. (20), we can calculate the $\tau_{em}$ as:

$$\tau_{em} = \frac{m^*}{e k_B T k_{avg}^3}\left(\mu_d T k_B k_{avg}^3 - 3\tau_m c^2 e p \pi^2\right) \qquad (24)$$

The $\tau_m$ can be calculated from eq. (1). From the Hall data and $\tau_m$, $\tau_{em}$ can be obtained from eq. (24) and we can also calculate $\tau_{me}$ using eq. (23). Using this approach, we estimated the three mentioned lifetimes for MnTe from the experimental data of the drag thermopower and Hall mobility, as illustrated in Figure 9.

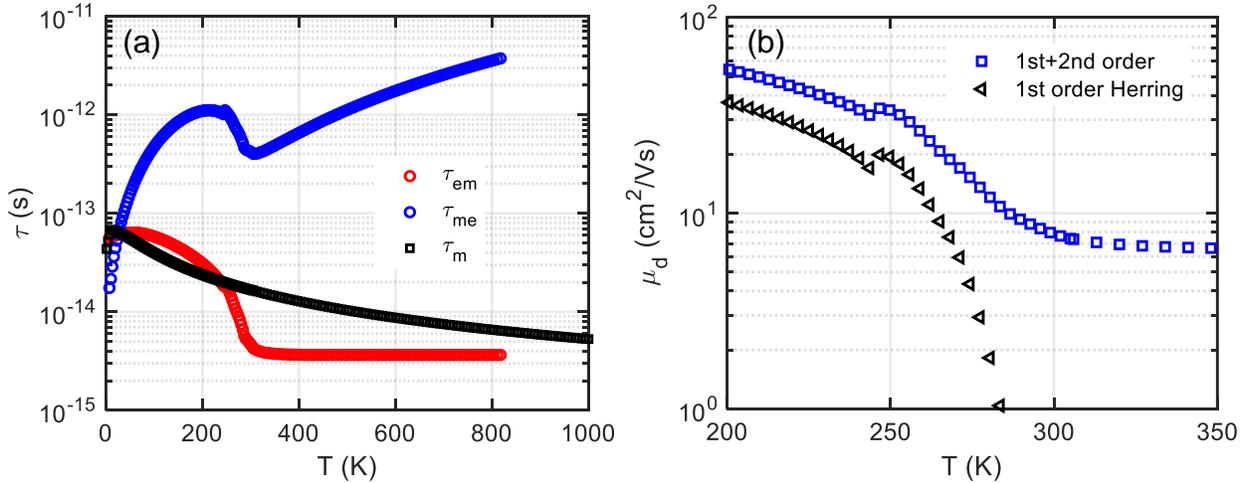

Figure 9: (a) Empirical estimations of the relaxation times $\tau_m$, $\tau_{em}$, and $\tau_{me}$ from eq. (1) and the Hall mobility data. (b) Comparison of the 1$^{st}$ order carrier mobility compared with the total (1$^{st}$ and 2$^{nd}$ orders) following Herring's model [81].

Here, the carrier mobility is assumed to be dominated by the magnon scattering, and all other scatterings are neglected. The nearest neighbor magnetic exchange energies $J_1$, $J_2$, $J_3$ (Figure *1*) are assumed to be -1.85 meV, +0.06 meV, and -0.25 meV, respectively, according to [74]. The positive and negative signs refer to FM and AFM exchange, respectively. The lattice constants are assumed to be a=4.144 Å and c=6.703 Å.

Figure 9 shows what the relaxations must be if the models describing eqs. (1), (20), and (23) are accurate. In Figure 9(b), we have calculated the first-order carrier mobility using the obtained lifetimes and compared it with the total carrier mobility [78,80]. It can be seen that the carrier mobility is significantly enhanced due to the second-order effect.

Several reasons can prevent the spin-disorder scattering theory from explaining the carrier mobility of the system:

1) The *s-d* exchange coupling between magnons and carrier system, which is the source of spin disorder scattering, is treated as a perturbation where energy uncertainty should be less than the thermal energy, $k_B T$, i.e.:

$$\Delta E = \frac{\hbar}{\tau} \ll k_B T \qquad (25)$$

   By replacing the $\tau$ with carrier mobility, the inequality relation in eq. (25) can be written as:

$$\mu \gg 45 \left(\frac{300}{T}\right) \; (cm^2/Vs) \quad (26)$$

   Here, eq. (26) assumed a carrier effective mass of $m_e$. According to the experimental values, the carrier mobility of MnTe violates the condition given in eq. (26).

2) The model assumes a quasielastic scattering where the energy transfer between magnon and electron systems is small compared to the thermal energy $k_B T$; hence, it does not apply to inelastic scattering. In degenerate FM semiconductors and metals, the quasielastic condition occurs at $T_c$ and above $T_c$; however, in AFMs, low-temperature optical magnons rise to inelastic scattering. This may not be critical for MnTe, a simple two-sublattice AFM with mainly two branches of magnon states of the acoustic type. However, the scattering process is inelastic for the temperature range where $k_B T < \hbar \omega_q$, for which the spin disorder scattering model cannot be applied. One can estimate the temperature at which the model fails by considering that magnons can scatter only a carrier of wave vector k with $q < 2k + mc/\hbar$. This scenario in AFMs is similar to the scattering of a carrier with acoustic phonons.[100] Therefore, for temperature range where $k_B T < 2\hbar ck + mc^2$, the electron magnon scattering is inelastic, and the model is not valid. For non-degenerate undoped MnTe, $k \cong k_{th}$, and for degenerate Li doped MnTe, $k \cong k_F$, where $k_{th}$ and $k_F$ correspond to the thermal and Fermi energies. Therefore, one can show that the inelasticity could be important for MnTe up to ~130 K and for $Mn_{0.97}TeLi_{0.03}$ up to ~500 K.

3) According to Haas [77], the contribution of the $q^2$ term is ignored, and only small $k' - k$ wavevector contributions are taken into account, while this assumption is not applicable for intervalley scattering. MnTe valance band consists of several valleys near the band extremum [101].

4) The spin relaxation time must be calculated from the complete wavevector spectrum; thus, its partial consideration can lead to a wrong estimation of the relaxation time.

5) The AFM order defines a new BZ boundary that can change the band energy and lead to effective mass variation at the Néel temperature.

6) Apart from the temperature-dependent effective mass, temperature-dependent wavefunctions can alter the exchange integrals, $J$, in the model calculations.

7) Moreover, the exchange integral can be different for the two MnTe sublattices due to different wavefunctions for the spins of opposite orientations.

The strength of these effects depends primarily on various parameters such as the band structure, magnetic exchange parameters, deformation potentials, temperature, etc. Therefore, quantitative analyses are required to determine which effects are dominant for a given material system.

In summary, as we discussed, thermoelectric materials progress based on the engineering of the electronic and phononic characteristics is reaching a plateau mainly because electrons are fermions, and the Fermi-Dirac statistics impose an inverse relation between the thermopower and the carrier concentration. Magnons and paramagnons are bosonic quasiparticles that can play as a new independent variable not limited to the counter-balancing nature of the parameters that enter $zT$. Recent studies in spin-driven thermoelectrics have demonstrated several auspicious spin-based effects, stimulating growing interests in magnetic thermoelectrics. Observations of thermopower enhancement and $zT$ improvement in the deep paramagnetic domain from entirely spin effects extend the search domain for good thermoelectrics to the paramagnetic semiconductors, of which there are many. Despite the considerably large landscape of magnetic semiconductors, thermoelectric material identification and design are challenged by a lack of reliable predictive tools to guide materials development. Several spin-based transport theories have been proposed to explain the current experimental data. We evaluated the successes and limitations of those theories for a case study on the MnTe system, a simple binary AFM semiconductor. The anomalous trend in carrier mobility, excess heat capacity contribution, and excess thermopower below and above the Néel temperature can be attributed to different spin effects. We especially present the applicability of several theories, such as spin-disorder scattering, magnon-drag, and paramagnon-drag effects. We also study the prospect of spin-entropy theory to explain the thermopower trend in the paramagnetic domain. Spin-fluctuation or spin-disorder scattering-based carrier mobility model fails to explain the experimental trend observed in MnTe. Significant $s$-$d$ interaction energy compared to thermal energy, significant inelastic processes, the quadratic magnon wave vector terms, improper consideration of wavevector spectrum, band energy, and BZ boundary are some of the factors that, as discussed, can lead to inaccurate estimation of the carrier mobility. Excess thermopower is attributed to magnon hole drag in the AFM regime consisting of both the first and second-order drag effects. The heat capacity-based models explain better the $T^3$ trend of the magnon-drag thermopower in the AFM domain but must be scaled significantly by an arbitrary coefficient to fit the data. The magnetic heat capacity of MnTe has contributions from both the spin-wave and magnetic ordering-disordering, although thermopower models based on heat capacity consider only the spin-wave component. In the PM domain, the anomalous thermopower is suggested to result from a paramagnon drag effect due to the nearly constant paramagnon lifetime observed from inelastic neutron scattering measurements. However, the extension of the current magnon-carrier drag theories to the paramagnon domain predicts a decline at the phase transition temperature, in contrast to the experimental observations. Furthermore, the experimental values are higher than the theoretical limits of the drag thermopower in the PM regime, meaning that either the existing theories are incomplete, or there are different or multiple spin-based mechanisms affecting the thermopower. The evaluation of the magnetic and spin entropy contributions to the paramagnetic thermopower shows that they could potentially explain the paramagnetic thermopower trend but with somewhat higher values. Overall, this work can explain the magnon and paramagnon drag thermopowers closer to the experimental values than the previous works. Further experimental and theoretical studies are needed to explain the anomalous thermoelectric properties precisely in magnetic semiconductors. The following guidelines can be drawn to help designing high performance spin-driven thermoelectric materials:

- AFM semiconductors are generally better candidates than FM ones for spin-driven thermoelectrics. AFM magnons have degenerate magnon modes, higher magnon group velocity, and longer magnon lifetime than FM magnons which provide enhanced first-order magnon-carrier drag thermopower for AFMs. Though, FMs give higher second-order drag thermopower than that of AFMs due to the longer magnon-electron relaxation time. Overall, drag thermopower is higher in AFMs than FMs due to the first-order drag effect's dominance.
- Thermopower is proportional to magnetic heat capacity. Electrical conductivity saturates above the transition temperature. Hence, materials with a significant magnetic heat capacity contribution that extends to temperatures above the phase transition can offer larger spin-driven thermopower leading to a high zT in the PM domain. The thermal conductivity increases with the heat capacity, but since the thermal diffusivity shows an opposite trend, thermal conductivity is less affected.
- Materials with a paramagnon lifetime higher than electron-on-magnon relaxation time offer more significant spin-driven thermopower.
- Thermopower increases linearly above phase transition, and spin disorder scattering saturates in the PM domain. Therefore, the high zT happens above and away from the phase transition temperature. This hints that the material's magnetic phase transition temperature must be well below the thermoelectric device's working temperature but not too low that the PM lifetime decays out.
- A high discrepancy of the exchange energies among the nearest neighbors in a crystal is likely helpful to sustain the local ordering in one direction (corresponding to the largest J) above the transition temperature. The phase transition happens when the magnetic ordering along the smaller exchange energies disappears, but a short-range correlation is still maintained along the largest exchange energy chain.
- There is an apparent drop in the MnTe paramagnon lifetime compared to the magnons at around $T_N$; however, the drag thermopower does not show any drop at the phase transition temperature. This observation is critical for designing spin-drive thermoelectrics, but the current theories cannot explain it.
- The magnon drag thermopower near and above $T_N$ is not a strong function of the charge carrier effective mass while the low-temperature thermopower is. Like the magnon-drag thermopower, magnon lifetime is also a weak function of the carrier effective masses in low mobility materials around $T_N$, which is valid for most magnetic materials.
- Magnon-drag thermopower starts showing the enhancement over diffusion thermopower roughly around the temperature $T_0$ at which the average magnon energy is equal to the thermal energy $k_B T_0/2$.

**Experimental Procedures**

**Resource availability**

*Lead contact*

Further information and requests for data should be directed to and will be fulfilled by the lead contact, Daryoosh Vashaee (dvashae@ncsu.edu).

*Materials availability*

This study did not generate new unique materials.

*Data and code availability*

All data related to this study included in the article and Supplemental information will be provided by the lead contact upon reasonable request.

**Sample Preparation**

Samples with the nominal compositions of $Mn_{1-x}Li_xTe$ (x=0, 0.03, and 0.05) were synthesized from raw elements (Mn powder, 99.99%, Li chunks, 99.9%, Te chunks, 99.999%). Samples are made inside argon-filled Tungsten-Carbide (WC) cups using a high-energy Fritsch P7PL planetary ball mill, keeping 5:1 WC balls to powder weight ratio. The materials were milled for 8hrs, annealed for 24hrs at around 1050K, milled again for 8hrs, and then sintered at 1173K for 20 min by spark plasma sintering (SPS) under an axial pressure of 50MPa with a heating rate of 50K/min. The ingots are cylindrically shaped with around 6.0 mm diameter, 12 mm length, and densities of >97% of theoretical values (6.0g/cm$^3$).

**Sample Characterizations**

The phase analysis was performed by X-ray diffraction (Rigaku MiniFlex, XRD) (See Figure S1). The resistivity and thermopower measurements were performed on samples simultaneously with the standard 4-point probe method using Linseis equipment under He environment. The commercial software of the equipment does not eliminate the dark emf and can lead to significant errors. Therefore, the thermopower was obtained from the linear fit to five separate temperature and voltage gradients, repeated four times for a total measurement of twenty points at each temperature. The accuracy of the measurement was confirmed by inspection. Thermal diffusivity ($\upsilon$) was performed on a thin disk (cut from the cylindrical ingot with a diameter of 6 mm, thickness <0.6mm) in the same direction as that of the electrical conductivity and thermopower using the laser flash apparatus (Linseis) under a vacuum environment from 250K-900K. The thermal conductivity ($\kappa$) was calculated from the relation $\kappa=\rho C_p \upsilon$, where mass density, $\rho$, is obtained by the Archimedes method and heat capacity, $C_p$, is obtained from PPMS and DSC. Low temperature (4K-400K) heat capacity is performed with Quantum Design 12T DynaCool physical property measurement system (PPMS), and high temperature (300K-900K) heat capacity is measured with Differential Scanning Calorimetry (DSC) under $N_2$ flow to avoid the formation of oxide phases. Low temperature (4K-400K) thermal transport properties, including electrical and thermal conductivity and thermopower, are also measured with PPMS.

**Supplemental Information**

Supplemental information can be found online.

**Acknowledgments**

The assistance of Douglas Abernathy and Rick Goyette in acquiring inelastic neutron scattering data at the ARCS, SNS, ORNL instruments is gratefully acknowledged. This study is partially based upon work supported by the Air Force Office of Scientific Research (AFOSR) under contract number FA9550-19-1-0363 and the National Science Foundation (NSF) under grant numbers ECCS-1711253 and CBET-2110603. Inelastic neutron scattering work at ORNL by DM, JZ, and RPH was supported by the U.S. Department of Energy (DOE), Office of Basic Energy Sciences, Materials Sciences, and Engineering Division. This research used resources at the Spallation Neutron Source and the Center for Nanophase Materials Sciences, DOE Office of Science User Facility operated by the Oak Ridge National Laboratory.

**Author Contributions**

D.V. directed the research. D.V. and M.H.P. performed the theoretical calculations and analyzed the results. M.H.P. synthesized the samples and characterized the transport properties. D.M., J..Z., and R.P.H. performed the neutron experiment and analyzed the data. All authors discussed the experimental results and contributed to prepare the manuscript.

**Declaration Of Interests**

The authors declare no competing interests.

# Supplementary Information

**Supplementary Experimental Procedures**

**XRD Characterization for Chemical Phase Analysis**

Manganese telluride, MnTe, an α-type AFM semiconductor, has NiAs hexagonal crystal structure with a=4.144 Å and c=6.703 Å. Due to the superexchange interaction between Mn 3d electrons via Te 5p electron (shown in Figure 2), MnTe shows AFM nature with Neel temperature of $T_N \approx$ 307K and PM Curie temperature of $\Theta=-585$K. MnTe has a direct bandgap of ~0.8 eV and an indirect bandgap of ~1.2 eV both below and above $T_N$ [76-78] with effective masses of $m_\perp = 0.4m_e$ and $m_\parallel = 1.6m_e$ [76-78]. MnTe shows hybridization of 5p Te and 3d↑ Mn bands in the valence band and hybridization of 4s Te and 3d↓ Mn band in conduction band [76-78]. The phase analysis was performed by X-ray diffraction (Rigaku MiniFlex, XRD) using Cu-Kα radiation and operating under 40 kV and 15 mA. XRD patterns of the samples are shown in Figure S1.

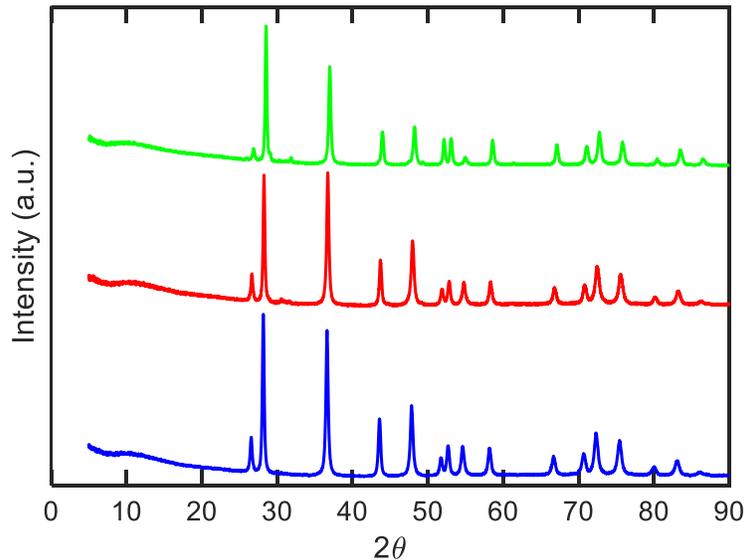

Figure S1: XRD patterns for synthesized $Mn_{1-x}Li_xTe$ samples (x=0, 0.03, and 0.05).

**Supplemental Table**

Different material properties are used in calculating the transport properties of MnTe. Table S1 represents the used parameters along with the corresponding references. Values without references are either taken from fitting experiments or are the known values.

Table S1: Different material properties of MnTe used in different numerical modeling

| Material Properties | Unit | Value | Ref./Remark |
|---|---|---|---|
| **Common For all numerical models** | | | |
| Density | g/cm$^3$ | 6 | |
| Neel Temperature ($T_N$) | K | 305 | |
| Paramagnetic Curie Temperature ($\theta$) | K | 585 | |
| Debye Temperature ($\theta_D$) | K | 120 | [68] |
| Einstein Temperature ($\theta_E$) | K | 180 | From fitting |
| Coefficient of Thermal Expansion, $\alpha_{CTE}$ | 1/K | 40×10$^{-6}$ | [68] |
| Gruneisen Parameter, $\Gamma$ | | 2 | [68] |
| Scattering Exponent, $r$ | | ½ | For s-d scattering |
| Effective mass, $m^*$ | kg | 9.1×10$^{-19}$ | |
| Spin Number, $S$ | | 5/2 | |
| g-factor, $g$ | | 2 | |
| Bandgap, $E_g$ | eV | 0.8 | |
| Magnon Velocity, $c$ | m/s | 14000 | [62] |
| **Exchange Energies:** | | | |
| $J_1$ | K | -21.5 | [65] |
| $J_2$ | K | 0.67 | [65] |
| $J_3$ | K | -2.87 | [65] |
| **Parameters used for Sugihara's Model** | | | |
| s-d Exchange Energy, $J_{sd}$ | eV | 0.2 | From fitting |
| Exchange Energy, $J$ | $k_B$ | 6 | [72,84] |

## Supplemental Figures

### Magnetic Susceptibility of Single Crystal MnTe

In spin-disorder scattering theory, spin-related relaxation times are obtained from magnetic susceptibility. Total magnetic susceptibility is considered as $(\chi^\parallel + 2\chi^\perp)$, where $\chi^\parallel$ is the susceptibility in the $c$ direction and $\chi^\perp$ is in the $a$-direction. To calculate the spin-flip and non-spin-flip scattering lifetimes, we adopted those data for single crystal MnTe from Komatsubara et al., J. Phys. Soc. Japan, 18(3), 356-364 (1963). The data was plotted in the SI unit (J/T$^2$kg) in Figure S2. We also measured the temperature-dependent magnetic susceptibility at a low excitation field (1000 Oe) for undoped and Li-doped MnTe, as shown in Figure S2(b). The trends observed in experimental magnetic susceptibility support the previously reported data. According to the figure, the MnTe samples are completely AFM up to the $T_N$.

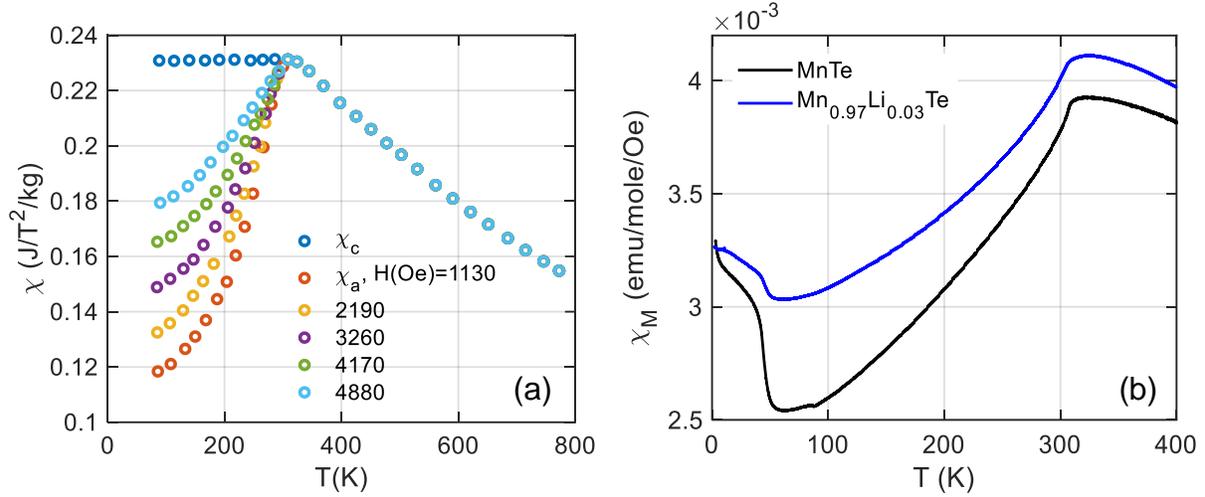

Figure S2: (a) Magnetic Susceptibility of single-crystal MnTe along a and c axis under various magnetization fields. Original data is adopted with permission from Komatsubara et al., J. Phys. Soc. Japan, 18(3), 356-364 (1963). (b) The magnetic susceptibilities of undoped and 3% doped polycrystalline MnTe.

**Relation between thermopower and resistivity**

The relation between thermopower and resistivity according to the Eq. 6 is shown in Figure S3.

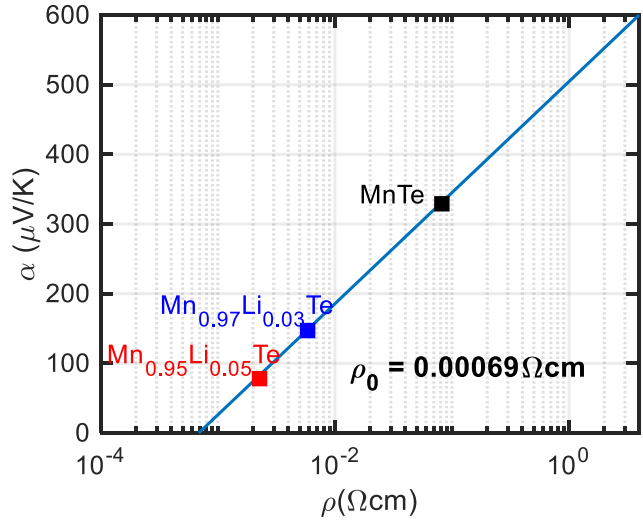

Figure S3: Thermopower of undoped and doped MnTe samples as a function of the resistivity.

**Supplemental Notes**

**Derivation of the relation between $\tau_{em}$ and $\tau_{me}$**

From the conservation of linear momentum, and considering only scattering of electrons and magnons, one can write [62]:

$$\frac{dK}{dt} = -\frac{dQ}{dt} = -\frac{K}{\tau_{em}} + \frac{Q}{\tau_{me}} \quad (a1)$$

Here K and Q are the electron and magnon system wavevectors, respectively. Therefore, at a steady-state, we have:

$$\frac{\tau_{em}}{\tau_{me}} = \frac{K}{Q} \qquad (a2)$$

Also, at steady-state, electrons and magnons must have similar drift velocities, i.e.:

$$v_d^e = v_d^m \qquad (a3)$$

We have for the electron system:

$$\hbar K = pm_e^* v_d^e \text{ or } v_d^e = \hbar K/pm^* \qquad (a4)$$

Here, $p$ is the charge carrier concentration, and $m^*$ is their effective mass.

For the magnon system, we have:

$$v_d^m = \frac{1}{\sum n_q} \left| \sum n_q \left(\frac{\mathbf{q}}{q}\right) v_m \right| = \frac{1}{\sum n_q} \frac{Q}{\bar{q}} v_m \qquad (a5)$$

Here, $v_m$ is the average velocity of magnons-, **q** is the magnon wavevector, and q its magnitude. $\sum n_q$ is the total number of magnons that participate in the scattering process. $\bar{q}$ is the average magnon wave vector.

The number of magnon modes in q space with $q \leq \bar{q}$ is:

$$M = 2 \frac{^4/_3 \pi \bar{q}^3}{8\pi^3} = \frac{\bar{q}^3}{3\pi^2} \qquad (a6)$$

The factor of 2 is to account for the magnon degeneracy of the AFM material. The number of magnons in a particular mode q with small q is $n_q = \frac{k_B T}{\hbar \omega_q}$. Therefore, the total number of magnons is:

$$\sum n_q = n_q M = \frac{\bar{q}^3}{6\pi^2} \frac{k_B T}{\hbar \omega_q} = \frac{\bar{q}^3}{3\pi^2} \frac{k_B T}{\hbar \bar{q} v_0} = \frac{k_B T \bar{q}^2}{3\pi^2 \hbar v_m} \qquad (a7)$$

We have from eq. (a5) and (a7):

$$v_d^m = \frac{3\pi^2 \hbar v_0^2 Q}{k_B T \bar{q}^3} \qquad (a8)$$

Also, from eq. (a4) and (a8):

$$\frac{3\pi^2 \hbar v_m^2 Q}{k_B T \bar{q}^3} = \frac{\hbar K}{pm_e^*} \Rightarrow \frac{K}{Q} = \frac{3\pi^2 pm^* v_m^2}{k_B T \bar{q}^3} \qquad (a9)$$

we have from eq. (a2) and (a9):

$$\frac{\tau_{me}}{\tau_{em}} = \frac{k_B T \bar{q}^3}{3\pi^2 pm^* v_m^2} \qquad (a101)$$

**Magnon carrier drag thermopower vs. temperature weighted mobility**

Eq. (7) of main text can be rearranged as:

$$\alpha_d = -\frac{k_B}{e} \log \left( 2e \left(\frac{600\pi m^* k_B}{\hbar^2}\right)^{3/2} e^{r+5/2} \rho_0 \mu_t \left(\frac{T}{300}\right)^{3/2} \right)$$

$$= -\frac{k_B}{e} \log \left( 2e \left(\frac{600\pi m^* k_B}{\hbar^2}\right)^{3/2} e^{r+5/2} \rho_0 x \right) \quad (a11)$$

Here, $x = \mu_t \left(\frac{T}{300}\right)^{3/2}$ is temperature-weighted mobility. For 100K condition, eq. (a11) is modified such that $= \mu_t \left(\frac{T}{100}\right)^{3/2}$. At T=100K, the experimental mobility is ~200 cm²/Vs; hence, x=100 in Figure S4. It can be seen that the drag thermopower becomes a strong function of the effective mass at this temperature.

We can further determine the magnon lifetime by combining eqs. (2) and (3) from the main text and (a11):

$$\alpha_d = \frac{\tau_m c^2}{\mu_{em} T} = -\frac{k_B}{e} \log \left( 2e \left(\frac{600\pi m^* k_B}{\hbar^2}\right)^{3/2} e^{r+5/2} \rho_0 x \right) \quad (a12)$$

$$\tau_m \frac{\mu_t}{\mu_{em}} = -\frac{300^{3/2} k_B}{ec^2 T^{1/2}} x \log \left( 2e \left(\frac{600\pi m^* k_B}{\hbar^2}\right)^{3/2} e^{r+5/2} \rho_0 x \right) \quad (a13)$$

From eq. (a13), we can plot the magnon lifetime as a function of $x$, illustrated in Figure S4 for $T = 100K$. Eq. (a13) can be modified similarly for 100K conditions.

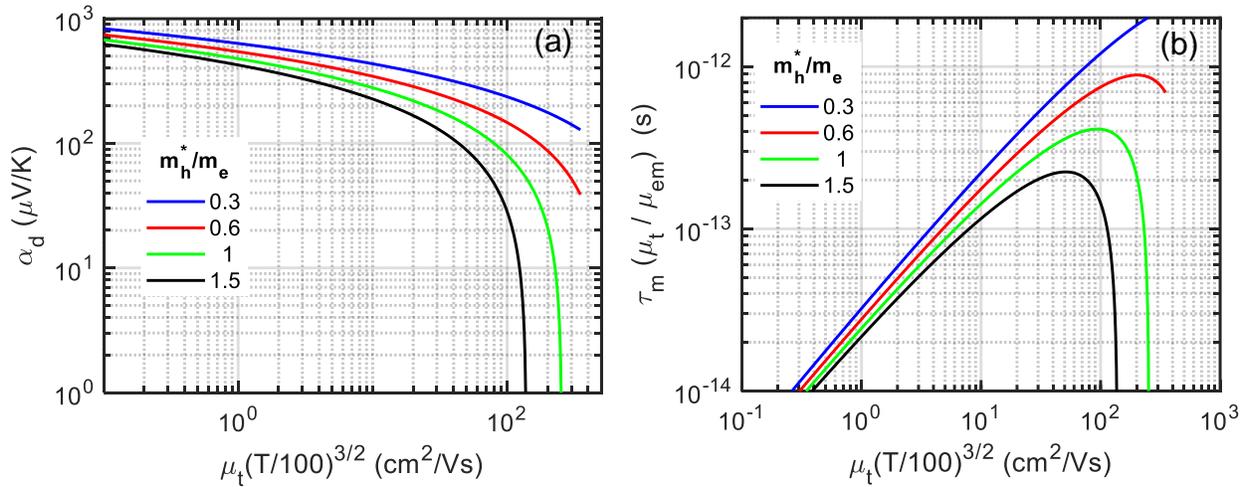

Figure S4: Magnon-drag thermopower (a) and magnon lifetime (b) as a function of temperature weighted mobility (*x*) for different hole effective masses at T=100K.

**Determination of Elastic and inelastic scattering temperatures**

The magnon-carrier scattering is similar to the case of electron scattering by acoustic phonons [79], where there is a limit to the wavevector of the magnons that can scatter an electron with wavevector k, i.e., $q \leq q_m = 2k + \frac{m^* v_m}{\hbar}$. This relation indicates that the energy of an electron can change after scattering by magnons at most by $\hbar v_m q_m = 2\hbar v_m k + m^* v_m^2$. Comparing this energy with $kT$, one can find the temperature $T_i$ below which the effect of inelastic scattering may become significant (similar to the case of Debye temperature) [80].

$$T_i = 2\hbar v_m + m^* v_m^2/k \qquad (a14)$$

We may check if the inelastic scattering is significant in MnTe. For the case of undoped MnTe, assuming $k \cong k_T$, one can find:

$$T_i \cong 10\, m^* v_m^2/k \qquad \text{(non-degenerate)} \qquad (a15)$$

Assuming $v_m \cong 1.4 \times 10^4$ m/s, and $m^* \cong m_e$, for the holes in MnTe, $T_i \cong 130$ K. For the degenerate case, $k \cong k_F$, and one has $q_m = 2k_F + m^* v_m/\hbar \cong 2k_F = 2(3\pi^2 p)^{1/3}$. Considering $kT_i = \hbar v_m q_m$, one can find:

$$T_i \cong 2\hbar v_m k_F/k \qquad \text{(degenerate)} \qquad (a16)$$

**Calculation of Magnetic heat capacity from inelastic neutron scattering**

To calculate the magnon heat capacity, we estimated the magnon density of states (DOS) from the INS spectra at 250K and 290K (shown in Figure S5). The heat capacity at low temperature can be estimated based on linear spin-wave approximation from the following relation:

$$C_m(T) = R \int_0^\infty \frac{x^2 e^x}{(e^x - 1)^2} g(\nu) d\nu \qquad (a17)$$

Here, $T$ is the temperature, $R$ is the ideal gas constant, $x = h\nu/k_B T$, $g$ is the magnon DOS, and $\nu$ is the magnon frequency.

The extracted magnon DOS was assumed to be constant at all temperatures. Therefore, the estimated magnon heat capacity can be off from the practical values, although its low-temperature trend must be similar to the illustrated one in Figure S5(b). Note that the leaner spin-wave approximation breaks near $T_N$, explaining why the calculated heat capacity is saturated near $T_N$.

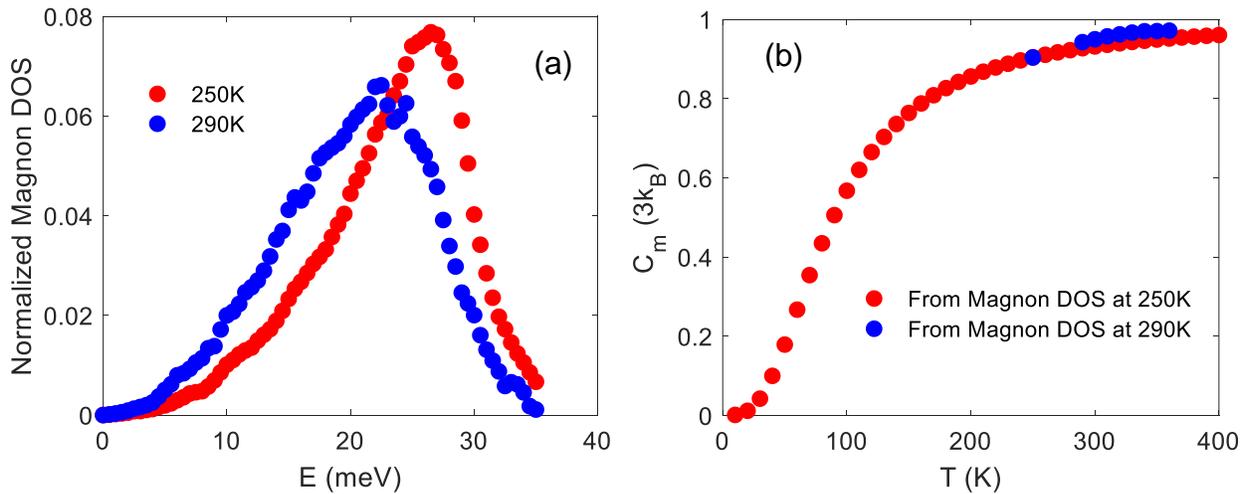

Figure S5: (a) Normalized magnon density of states (DOS) obtained from the inelastic neutron scattering data at 250K and 290K, and (b) calculated magnon heat capacity from the estimated magnon DOS. The estimated magnon heat capacity based on linear spin-wave approximation is valid only at low temperatures and can be significantly off near the transition temperature (~310K for MnTe).

## Calculation of Spin-flip and spin non-flip scattering lifetime

Both energy-dependent scattering mechanisms, namely spin-flip and spin non-flip scattering, can be determined from magnetic susceptibility and magnon band structure based on the relation given by eqs. (21) and (22). The calculated spin scattering lifetimes are illustrated in Figure S6.

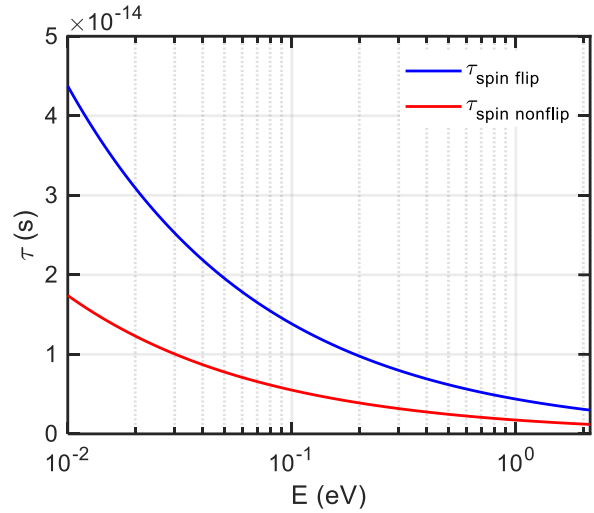

Figure S6: Energy-dependent spin-flip and spin non-flip scattering lifetimes at 250K, assuming a hole concentration of $10^{19}$ cm$^{-3}$.

## Field-dependent thermopower, magnetic moment, and heat capacity

Magnetothermopower is measured to investigate the impact of an external field on spin entropy thermopower, as illustrated in Figure S7(a). Figure S7(b) and (c) illustrate the field-dependent magnetic moment and heat capacity. The data are discussed in the manuscript.

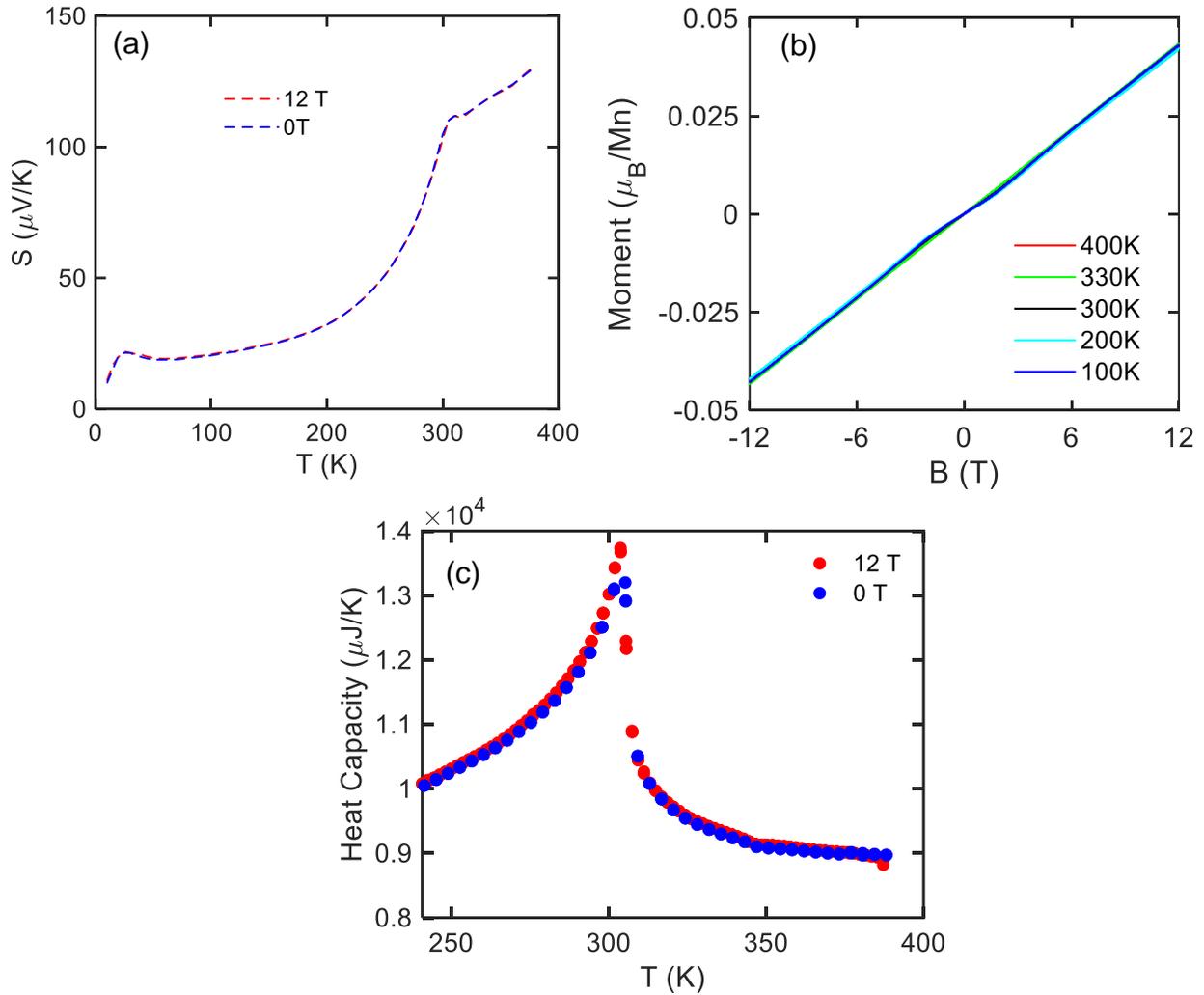

Figure S7: (a) Field-dependent thermopower at 0T and 12T, (b) field-dependent magnetic moment at different temperatures, and (c) field-dependent heat capacity of Li-doped MnTe.

**Field-dependent Hall Data Analysis**

Field-dependent Hall data is measured and analyzed to calculate carrier mobility. Figure S8 shows the field-dependent electrical conductivity and Hall resistivity data of MnTe at selective temperatures from magnetic and paramagnetic domains.

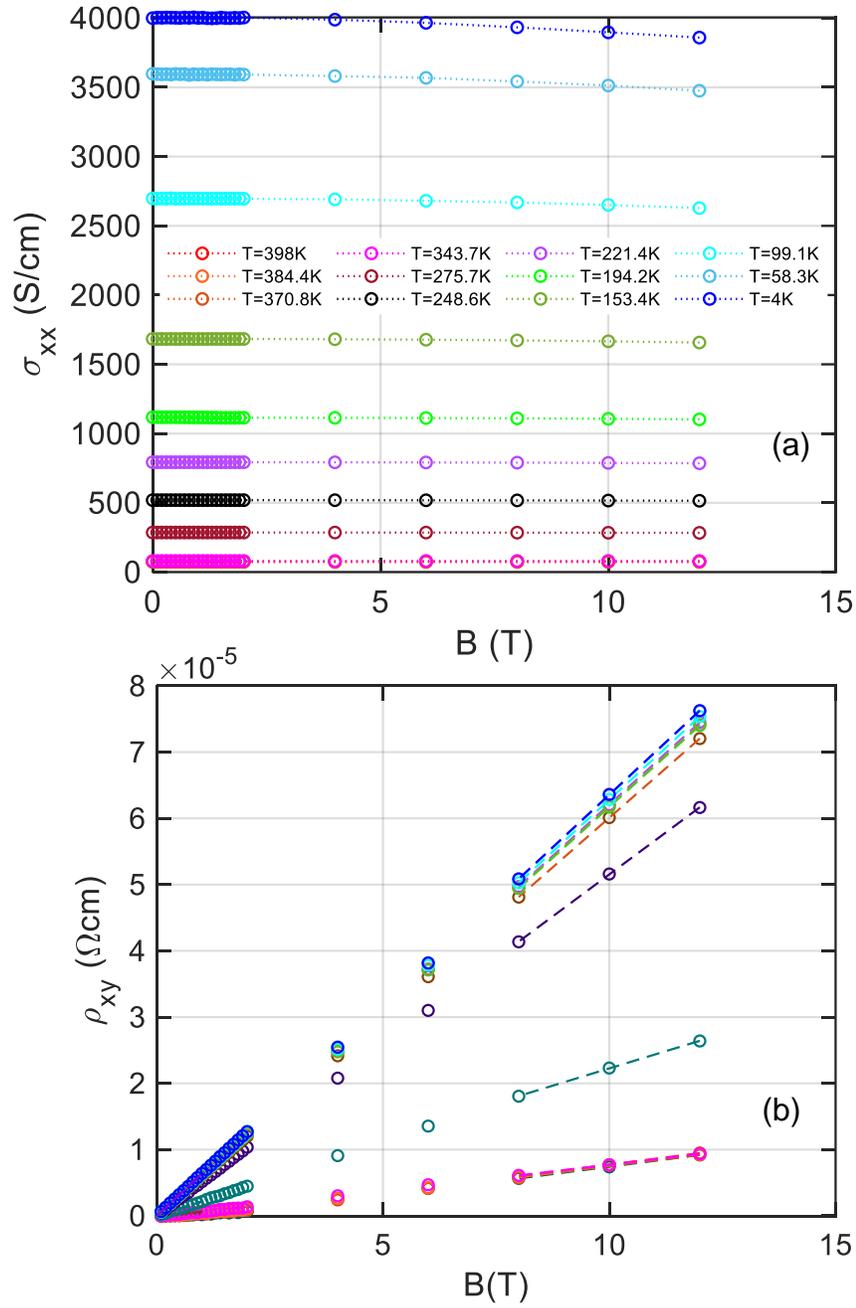

Figure S8: (a) Field-dependent electrical conductivity and (b) field-dependent Hall resistivity of MnTe at different temperatures from magnetic and paramagnetic phases for MnTe.